VICTOR VAN REIJSWOUD
ARJAN DE JAGER

# FREE AND OPEN SOURCE SOFTWARE FOR DEVELOPMENT

exploring expectations, achievements and the future

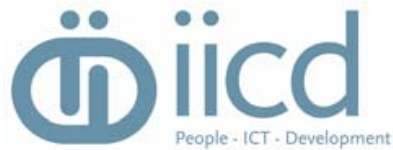

Polimetrica

PUBLISHING STUDIES

directed by Giandomenico Sica

VOLUME 5

**VICTOR VAN REIJSWOUD**

**ARJAN DE JAGER**

# FREE AND OPEN SOURCE SOFTWARE FOR DEVELOPMENT

**exploring expectations, achievements and the future**

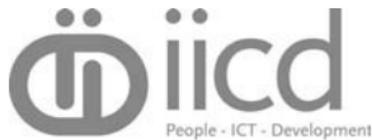

Polimetrica

**Note for the Reader**

In our view, doing research means building new knowledge, setting new questions, trying to find new answers, assembling and dismantling frames of interpretation of reality.

**Do you want to participate actively in our research activities?**

**Submit new questions!**

Send an email to the address **questions@polimetrica.org** and include in the message your list of questions related to the subject of this book.

Your questions can be published in the next edition of the book, together with the author's answers.

**Please do it.**

**This operation only takes you a few minutes but it is very important for us, in order to develop the contents of this research.**

Thank you very much for your help and cooperation!

We're open to discuss further collaborations and proposals.
If you have any idea, please contact us at the following address:

*Editorial office*
*POLIMETRICA*
*Corso Milano 26*
*20052 Monza MI Italy*
*Phone: ++39.039.2301829*
*E-mail: info@polimetrica.org*

**We are looking forward to getting in touch with you.**

"The box said that I needed to have Windows 98 or better...
so I installed Linux."
--- CARUS M. (221556)

There, I've said it. I'm out of the closet. So bring it on...
--- Linus Torvalds

Quotes on: http://www.ao.com/~regan/quotes/Linux.html

# LIST OF QUESTIONS

# INTRODUCTION

In 1991 Linus Torvalds used a new paradigm in software development that is now maturing and has the potential to change the world. Torvalds developed an operating systems called Linux. Initially he was interested in developing a small version of the UNIX operating systems. In order to improve the software he decided to share the code with the software community outside the University of Helsinki in Finland. The software community based approach in the development of Linux gave the real boost to the Free and Open Source Software (FOSS[1]) philosophy, since it was proved that it was able to produce software that was able to compete with commercially produced softwares (www.linux.org). The launch of the first Linux distribution (a combination of the operating systems and supporting applications) by Torvalds in 1994 has lead to an explosion new Linux based Open Source operating systems and application software to run on the Linux platform. At the moment of writing www.linux.org lists 220 different (maintained) Linux distributions.[2]

The FOSS philosophy challenges the general accepted software development paradigms that are used by companies of today (Raymond, 1998). Traditional software development paradigms are based on the idea that software has to be fully developed and tested before it is sold in the market. When the software is put in the market, users can not change the source code, and mistakes have to fixed by the software company. This way of working makes the development of new software a labor intensive and long process. With the development of Open Source Software, a different route is taken. The basic functionality is programmed by the initiator(s) and then made available for others to test, use and/or modify. Mistakes in the software are not considered problematic, but are accepted. Since the source code is distributed, every software engineer can change or extend the original product. So, where propriety software is developed in-house and then released, FOSS is under constant development because anyone in the world can change the code.[3]

An important aspect in pro-FOSS discussions is the price. Not all FOSS is distributed free of charge, and some come with a price tag, but in most cases it is cheaper to acquire than proprietary software. The real price difference emerges from the fact that there not a license fee structure. Where for proprietary software all the users need to pay a fee, in the FOSS approach someone buys the software, and becomes the owner and can start to freely redistribute it to other users. Especially in larger organizations this can make a huge difference.

Although a lot has been written about the importance of FOSS, its advantages and challenges, most is published in the context of the developed countries: Europe and the North America. Growing attention is noticed for the strong developing economies in Latin America, like Brazil, the Indian Subcontinent, India, and there is a strong promotion by the Asian-Pacific Development Information Programme (APDIP) for the use of FOSS in the countries in South East Asia. On the contrary, surprisingly little has been published on the use of FOSS on the African continent. Donors have promoted the use of FOSS since huge advantages are expected, projects have been funded, but the actual impact has not been well mapped.

This book is about FOSS for Development (FOSS4D). We will focus on the Least Developed Countries (LDC's) and primarily on the African context. Most of the LDC's are in Africa. Both authors have worked in this context and initiated and managed FOSS4D projects in several parts of Africa. It is on these experiences that we will build and expand. We are both convinced that FOSS can make a huge difference for the lives of the people and can greatly expand their access to information. FOSS will take away the financial and legal barriers that limit the use of software in schools, universities, civil society and at government levels.

The book will guide the reader to a better understanding of the role of FOSS for the development of the LDC's through a range of questions. The questions are related but provide answers in themselves. The reader is encouraged to read the questions in sequential order, but for readers that understand the potential of FOSS, the individual answers will help to make their position stronger. The examples that are used in the book are mostly based on the projects that are supported by the International Institute for Communication and Development (IICD) but they are not limited to the work of this organization.

Finally, this book is mainly based on Free and Open Content that has been made available through the internet or otherwise. We have refrained as much as possible from using Paid and Closed Content as a matter of principle. We believe that free and open exchange of knowledge is necessary for the development of LDC's and opening up content to limited groups of people (i.c. those who can afford) should be discouraged. We realize that this position may (not necessarily) limit the range of the book, but at the same time it makes all the underlying knowledge available and accessible for all readers.

Victor van Reijswoud
victor.vanreijswoud@gmail.com
Arjan de Jager
a.de.jager@hec.nl

## 1. What is the role of technology for the Least Developed Countries?

Most of the economies in the Least Developed Countries[4] (LDC's) are still agricultural economies that try to rush into the information age. This requires a rapid adoption of all kinds of technologies.

Information and Communication Technologies are a relatively recent instrument in the fight to eliminate hunger and poverty and increase the quality of life of the people living in the LDC's (Blommestein at al., 2006). The World Bank in its 2002 Strategy Paper on ICT states that:

*"Information and Communication Technologies are a key input for economic development and growth. They offer opportunities for global integration while retaining the identity of the traditional societies. ICT can increase the economic and social well-being of poor people, and can empower individuals and communities. Finally ICT can enhance the effectiveness, efficiency and transparency of the public sector, including the delivery of social services."* (World Bank, 2002)

ICT4D projects have been implemented in several sectors in the LDC's and gradually it becomes clear that successes are possible with ICT, but that the programs need to be designed and implemented with care. Early enthusiasm and claims that ICT would prove a silver bullet for development problems lead to a number of false starts. Many of the problems in the early period are to be blamed on the lack of experience of the project managers from both the donor countries as well as on the recipients side and the fact that solutions that worked in developed countries were unthinkingly copied to projects in LDC's. Over time the program managers have matured and the uniqueness of ICT solutions for LDC's is gradually recognized. The last is still underway and this book tries to contribute to this domain of knowledge. We consider Free and Open Source Software one of the solutions that may help LDC's to leap into the information age.

Not all people are convinced that ICT can contribute to an increased quality of the lives of the people in the LDC's. There are more important issues to be addressed, critics say. Daly (2003) puts the point clearly:

*"In a fundamental way, ICT's are not going to help these kids. They can't eat computers, telephones won't make them well. However, given people, policies and institutions working to solve the problems of hunger and malnutrition, ICT can make a difference."*

We do not promote that ICT presents a silver bullet for all the problems that the LDC's face, but it may provide them with access to the basic information and tools to make informed decisions that will trigger new levels of development.

## *2. What is the digital divide?*

**KEYWORDS:**
DIGITAL DIVIDE, ICT GAP, KNOWLEDGE DIVIDE, ACCESS TO ICT

Regardless of how we measure it, there is an immense information and communication technology (ICT) gap, a "digital divide", between developed and developing countries. Some statistics published by the ITU quantify some aspects of the digital divide.[5] In 2004:
- the developing world had 4 times fewer mobile subscribers per 100 people than the developed world;

- the developed world still had 8 times (was 73 in 1994) the Internet user penetration rate of the developing world;
- less than 3 out of every 100 Africans use the Internet, compared with an average of 1 out of every 2 inhabitants of the G8 countries (Canada, France, Germany, Italy, Japan, Russia, the UK and the US);
- there are roughly around the same total number of Internet users in the G8 countries as in the whole rest of the world combined: 429 million Internet users in G8 and 444 million Internet users in non-G8;
- the G8 countries are home to just 15% of the world's population – but almost 50% of the world's total Internet users;
- Africa accounted for 13% of the world's population, but for only 3.7% of all fixed and mobile subscribers worldwide;
- the top 20 countries in terms of Internet bandwidth are home to roughly 80% of all Internet users worldwide;
- there are more than 8 times as many Internet users in the US than on the entire African continent.

Relative to income, the cost of Internet access in a low-income country is 150 times the cost of a comparable service in a high-income country. There are similar divides within individual countries. ICT is often non-existent in poor and rural areas of developing countries (United Nations, 2006). This is partly due to the lack of infrastructure but another reason is the relatively high costs: Even when the costs are the same in both urban and rural areas, income disparities between rural and urban communities make communication services more expensive for rural dwellers. Within region in the LDC's there are also significant differences. Table 1 provides a detailed breakdown of the computer and internet usage in different areas in the world.

|  | **Computer Use (per 100 people)** | **Internet Use (per 100 people)** |
|---|---|---|
| Developing Countries | 2.5 | 2.6 |
| Least Developed Countries | 0.3 | 0.2 |
| Arab States | 2.1 | 1.6 |
| East Asia and the Pacific | 3.3 | 4.1 |
| Latin America and the Caribbean | 5.9 | 4.9 |
| South Asia | 0.8 | 0.6 |
| Sub-Saharan Africa | 1.2 | 0.8 |
| Central & Eastern Europe & CIS | 5.5 | 4.3 |
| OECD | 36.3 | 33.2 |
| High-income OECD | 43.7 | 40 |

Table 1: Computer and internet use in different regions (UNDP, 2006).

There are many definitions of the digital divide and although they differ slightly, they focus on the access to information and communication technology (telephones, computers and internet) and the skills people need to access information and knowledge that will increase the quality of their lives (Sciadas, 2003). Access is determined by many variables at national, community and individual levels. Some countries in the developing world have such a poor electricity and internet infrastructure causing computers and internet to be basically only available in the capital (Best et al., 2007). In some countries access to internet is so expensive that only the top-earners can afford it (Sciadas, 2003) and there are even governments that prefer to limit their citizens in accessing information on the internet. But even when people have access to ICT and internet, they still need to have to skills to use these technologies. Knowing how to switch a computer on and off does not automatically guarantee an entrance into the world of knowledge. It is important that users of ICT know how to use the computer to write a letter to their representative in parliament, or search and process information on the internet that will help them to prevent their crops from being eaten by locust. This last is also sometimes referred to as the knowledge divide. It is the combination of access and skills (or better the lack thereof) that will determine the magnitude of the digital divide.

A special dimension to the digital divide is presented by the information that is available on the internet. Most of the information that one finds on the internet is produced by the developed countries in the North (Europe and North America). According to the Internet World Stats[6], the top ten languages on the Internet, listed below, account for 81.8% of all Internet use. English is the dominant language, accounting for almost 30% of Internet users, with Chinese coming up. The LDC's have contributed considerably less to the public information domain. This is partly caused by their limited access to ICT and partly by the fact that potential contributors lack the necessary skills to add information. A result of this is that finding information that can be useful to the lives of the people and that can directly increase the quality of their lives is more difficult. The production of 'local' knowledge has been promoted strongly by initiative like the Development Gateway[7] and the Drumbeat[8], but there is still a long way to go.

Free and Open Source Software and Open Content can play an important role in bridging the digital divide. FOSS lowers the barriers for people to have access to tools that will enable them to access information and contribute information to the public domain and Open Content removes the barriers that the publishers of information put up to distinguish people that can afford to pay for information from the ones that cannot. We will develop this issue further in the course of this book.

## 3. How does access to information relate to development? An example

**KEYWORDS:**
ICT4D, RURAL FARMER COMMUNITIES, UGANDA COMMODITY EXCHANGE, ICT4D CASE STUDY

The role of access to information for the strengthening of communities can be argued from a theoretical perspective, but it can better illustrated with an example. Below we will present a project that is being implemented in Uganda: The Uganda Commodity Exchange (UCE). The description of the UCE is based on (Blommestein et al) Case: Combined Warehouse and ICT-assisted commodity trading in Uganda

*A truck loaded with three tons of coffee rocks towards a big and empty warehouse in Kabwohe, in Sheema district in southwest Uganda. It is the first delivery coming from three farmer societies. Instead of selling their coffee by means of the middleman, the goods stay here till the farmers agree on a price with the highest bidder on the electronic trading floor of the Ugandan Commodity Exchange (UCE) in Kampala.*

*Uganda is one of the poorest countries in the world. In 2004, at the beginning of the project, the per capita income was estimated to be approximately US$250. Life expectancy at birth remains low: 43 years in 2002, compared to 47 years in 1990. Similarly, infant and child mortality has not improved much over the same period and today remains at around 100 respectively 150 per 1,000 live births.*

*Data on the increase in agricultural production is hard to obtain but it certain that the increase in agricultural production is not keeping pace with the growth in population. The Uganda Commodity Exchange project addresses this issue as it aims to:*

- *Establish an efficient communication system to enable effective collaboration between all stakeholders in the agricultural sector*
- *Provide accurate and timely information from all sections of the agri-industry system*
- *Enable rural farmer groups to produce and trade in a more commercial manner*

*The Uganda Commodity Exchange was first established in 1998 and acts like a stock exchange through the auctions of agricultural commodities. In 2004 an information system (IS) was implemented to support the farmer groups. The UCE-IS informs farmers on a whole host of issues such as current prices, market trends, and price fluctuations, is critical as this enables them to make informed decisions with regard to production planning and pricing. The price/market information is collected by the farmer groups, shared, and disseminated using a variety of different media: announcements posted at the centers, team leaders linked to farmer groups who distribute the information to the farmers (traveling from group to group on bicycles or motorcycles), radio and SMS messages. At this stage the project reaches three centers with 24 farmer groups, each group with over 200 farmers, for a total of approximately 4800 farmers.*

*Analysis of the project impact showed an high increase of awareness among the farmers about the price fluctuations and role of the middlemen in the pricing structure of the*

*commodities. There are also strong indicators that production has increased and diversified. One of the participants in the project stated: "With better prices, our standard of living will improve and we shall even improve further the quality of coffee. Later we hope to sell beans, peas and honey in this way. Everybody will benefit".*

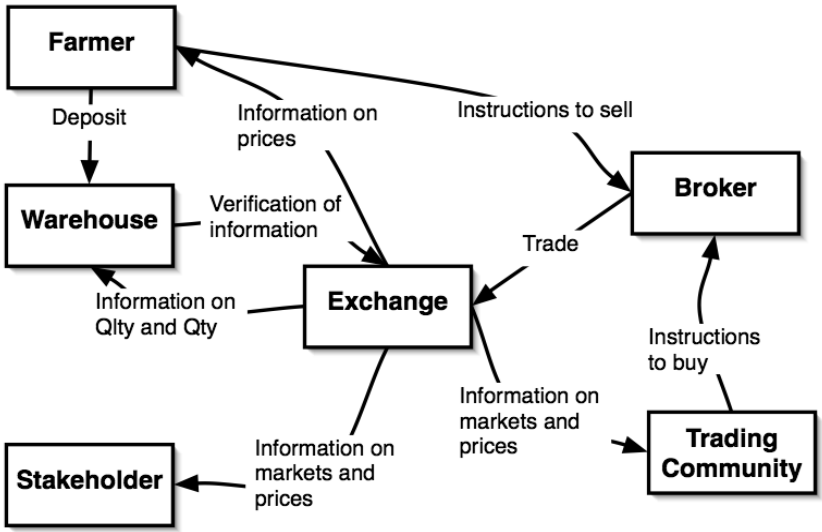

Figure 1: Information Flow Diagram at the Uganda Commodity Exchange.

Evidence shows that if and when farmers are able to access relevant and qualitative information regarding their production methods and commodities, they are able to increase their production levels as well as obtain better prices for their products. This benefits both the farmers and their families as well as the national economy. ICT support this in a variety of ways (Blommestein et al., 2006):

- Providing general information
- Access to new markets

- Empowering farmers to negotiate better prices
- Enhancing position in the value chain
- Optimizing usage and preservation of natural resources
- Support improved (financial) management processes

## *4. What are the major challenges for organizations LDC's implementing ICT4D?*

**KEYWORDS:**
LDC'S, BRIDGING THE DIGITAL DIVIDE, CHALLENGES FOR ICT4D, CAPACITY CHALLENGES, FINANCIAL CHALLENGES

Although ICT is an important tool to bridge the digital divide, the technology also brings along huge challenges for organizations in LDC's. These challenges can be divided into two main categories:

- capacity challenges
- financial challenges

We will address both challenges below.

**Capacity challenges**

ICT brought new and powerful technology for all LDC's. Where developed countries had already a relatively long history in which ICT has gradually been developed and integrated in the daily and organizational reality, LDC's were only confronted with it in the last 10-15 years, depending on the countries. Some countries like Kenya, Senegal or Zimbabwe had some limited experiences with ICT for some time, but countries with lower development levels, like Chad, the Democratic Republic of Congo, or the Central African Republic have virtually no experience with ICT dating before the introduction of donor-supported projects.

The consequence of this late introduction is that there was no or very limited knowledge infrastructure to support the use of ICT. Primary and secondary education is not providing basic computer-literacy programs, universities had no programs in computer science or information systems (or outdated and theoretical ones), decision-makers were not aware of the possibilities that the new technology was to offer, there were no trained business support and so on. In other words, the powerful technology landed in a knowledge and capacity vacuum. Expensive foreign experts were more than happy to fill in this vacuum.

In order to bring down the costs of development, implementation and maintenance of the ICT infrastructure, capacity needed to be build rapidly and with the right knowledge and skills. 'Old school' university curricula had to be replaced with programs that provide practical skills to students in order to be able to play an active role in the ICT development in the country. In most countries this process is still underway. Universities are gradually changing the programs and vocational training is offered for sub-university level students. Programs like the CISCO academy program for LDC's are important initiatives to improve the knowledge and skills levels to the required level.

**Financial challenges**

The introduction of ICT also brought financial challenges to those organizations eager to adopt the new technology. Next to the costs of training and educating people, as we have seen in previous section, acquiring hardware, ICT governance and software also poses challenges.

Computer hardware is often a large expense for organizations in the developing world, when compared to available financial resources. The costs of a simple computer (with internet connection and the necessary surge protection) are often comparable to the annual salary of the person using it.[9] The introduction of ICT, for example in a ministry in a developing country is accountable for a huge investment, which is in a lot of cases not available.

Computer software is an often forgotten and underestimated cost. Ghosh (2003) shows that what the developed world considers minor costs for productivity software like Microsoft Windows and Microsoft Office, becomes an exorbitant cost when it is related to the Gross Domestic Product of the LDC's. In figure 2 the price of Windows XP is expressed in the GDP Months for several countries and regions in the world. Prices of commercial software like databases, learning management systems, document management systems, software development environments etc. extend the costs of the ICT far beyond the investment costs of the hardware.

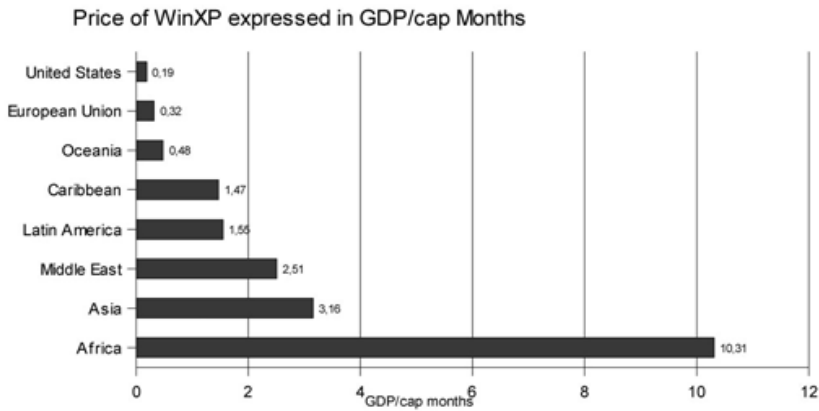

Figure 2: Price of Windows XP expressed in GDP/capita months (based on data in Ghosh, 2003).

Increased personnel costs are the last financial challenge that we would like to highlight. The introduction of ICT in organization is always accompanied by new internal or external staff members providing ICT maintenance and user support. Users needs to be trained, day-to-day problems will have to be addressed and solved, server and other systems will have to be maintained and updated and important information will have to be stored and protected. Soon after the first computers are introduced an ICT department is established. At national levels, the introduction of ICT may lead to new governing and regulating bodies, and increasing to the establishment of ICT ministries. These should be all considered ICT related costs.

## 5. What is the role of the donor community in promoting ICT4D?

**KEYWORDS:**
DONOR PROJECTS, DONOR COMMUNITY, ICT4D PROJECTS, DONOR RESPONSIBILITIES IN ICT4D

The donor community has an important role to play in promoting the use of ICT for development in the LDC's. In general terms, the donor community needs to guide communities in the LDC's in *discovering* the added value of ICT in improving the quality of life .

ICT has a wide range of application areas and the first role of the donor community is to play a guiding role for communities that want to explore the possibilities of ICT. The ICT revolution has brought about enormous changes in private and public spheres in the developed countries. Access and storage of information has never been greater and new information sharing and communication possibilities have really created a global village. For the developed nations this has been a gradual change, but the LDC's are confronted with a almost impenetrable range of possibilities. Many new entrants are paralyzed by these overwhelming possibilities.

They do not know what and where to search on the internet and they do not have an informed idea about the range of tools that they can use. Usually they end up using the internet for leisure and the computer as a sophisticated (but very expensive) typewriter. The donor community has a responsibility to guide LDC's as novices into a new world to make sure that they are not lost and that the investments are used in an efficient and effective manner.

This guiding role needs to be geared to discovery of needs and answers. When confronted with the new world of information, communication and technology, people first need to be aware of their needs and desired improvement in the quality of their lives. This will range from easier communication means, to access to information about the price of commodities as we have seen before. At a national level improved record-keeping of key economic indicators may be a key need. Whatever it is, the donor will have to provide assistance to governments, communities and individuals in revealing and articulating needs and prioritizing them. Only when the needs are clear the appropriate technologies can be selected.

Too often the donor community limits its role to a mere financial funder of the ICT infrastructure. They make available the financial means for the implementation of the ICT infrastructure and forget that they have to be a guide to make the potential beneficiaries discover the potential of ICT. This approach has resulted in many so-called 'white elephants', i.e., ICT infrastructures that are not used or fail to contribute to an improved quality of the life in the communities they are meant for.

The donor community will have to be serious about the importance of its role. ICT4D requires specialists that understand the context and have a good overview of the possibilities that are suitable for LDC's. Hardware that does not last in tropical and/or dusty conditions[10] or software that requires an online activation through a credit-card payment are typical examples of solutions that are not suitable to be used in ICT4D. Open and low-cost solutions will have to be in the toolkit that is presented by the ICT expert of the donors to the governments, communities and individuals in the LDC's.

## *6. What is Free and Open Source Software (FOSS)?*

**KEYWORDS:**
FREE AND OPEN SOURCE SOFTWARE (FOSS), ORIGINS OF FOSS, PROPRIETARY SOFTWARE, BILL GATES, RICHARD STALLMAN, FREE SOFTWARE FOUNDATION (FSF), GNU, OPEN SOURCE INITIATIVE (OSI)

*"Briefly, OSS/FS programs are programs whose licenses give users the freedom to run the program for any purpose, to study and modify the program, and to redistribute copies of either the original or modified program (without having to pay royalties to previous developers)." (Wheeler, 2003)*

Finding an agreement on one definition of Free and Open Source Software has proved to be difficult, but the definition of David Wheeler provides a good description of the essence of what FOSS is. It is software that is produced and issued by a community that likes to have their products open and likes them to be shared freely with the others in the community. It argues from the idea of a community that likes to learn and share without leaving people out. The FOSS community promotes the growth of knowledge by allowing other members to stand on the shoulders of the giants in this same community.

At the philosophical level we find two major schools or paradigms in the FOSS world: the oldest is the philosophy of the Free Software Foundation (FSF) philosophy founded by Richard Stallman. On the other end is the more business-like approach expressed in the Open Source Initiative (OSI) philosophy.

The Free Software Foundation has a long history rooted in the academic principles of knowledge sharing. The FSF emerged in the early days of computer science and computer industry when sharing software code became a problem and software gradually became 'closed'. Before this period software was treated as most academic products. People were sharing computer code, algorithms or whole programs with their peers. This sharing was done on the basis that you could use it, but had to acknowledge the origin of the information, the same way most of the academic world is still functioning.

The rise of industry and the commercialization of the computing industry changed this attitude. Sharing was gradually replaced by protection and academics that promoted openness had to make way for entrepreneurs that build 'closed'/proprietary software. By many, William (Bill) H. Gates' now-famous pamphlet: "An Open Letter to Hobbyists" dated 3rd February 1976, is considered a landmark in this change. In this letter Bill Gates rails against the prevailing culture of software sharing:

*"Why is this? As the majority of hobbyists must be aware, most of you steal your software. Hardware must be paid for, but software is something to share. Who cares if the people who worked on it get paid?"*

The gradual destruction of the software sharing culture Gates refers to was reason for Richard Stallman, researcher at MIT Artificial Intelligence Lab to stand up and promote the Free and Open Source Software development and licensing. He founded the Free Software Foundation.

According to the FSF, free software is about protecting four user freedoms:
- The freedom to run a program, for any purpose.

- The freedom to study how a program works and adapt it to a person's needs.
- Access to the source code is a precondition for this.
- The freedom to redistribute copies so that you can help your neighbor.
- The freedom to improve a program and release your improvements to the public, so that the whole community benefits. Access to the source code is a precondition for this.

At the heart of FSF is the freedom to cooperate and collaborate. Because non-free (free as in freedom, not price) software restricts the freedom to cooperate, FSF considers proprietary software unethical. FSF is also opposed to software patents and additional restrictions to existing copyright laws. All of these restrict the four basic user freedoms listed above.[11]

At the same time the world and the FOSS community is changing. Free and Open Source Software (FOSS) has become an international phenomenon, moving away from relative obscurity to being the basis of a full blown industry. Within the context of the approach of the FSF, business initiatives do not always feel comfortable. The approach of the Open Source Initiative likes to accommodate this. In the nineties, this group associated with FSF introduced the term "open source" to emphasize a break with the pro-hacker, anti-business past associated with GNU and other free software projects and to place a new emphasis in the community on the possibilities of extending the free software model to the commercial world. The new "open source" projects exist in the mainstream of the commercial software market and include operating systems, such as Linux, the Apache web server, and the Mozilla browser.

The OSI philosophy is therefore somewhat different from the FSF philosophy:

> *"The basic idea behind open source is very simple: When programmers can read, redistribute, and modify the source code for a piece of software, the software evolves. People improve it, people adapt it, people fix bugs. And this can happen at a speed that, if one is used to the slow pace of conventional software development, seems astonishing."*
> *(Wong, Sayo, 2003)*

The OSI is focused on the technical values of making powerful, reliable software, and is therefore more business-friendly than the FSF. It is less focused on the moral issues of Free Software and more on the practical advantages of the FOSS distributed development method. 998, a group associated with free software introduced the term "open source" to emphasize a break with the pro-hacker, anti-business past associated with GNU and other free software projects and to place a new emphasis in the community on the possibilities of extending the free software model to the commercial world. These new "open source" projects would exist in the mainstream of the commercial software market and include operating systems, such as Linux, the Apache web server, and the Mozilla.

OSI defines Open Source as software providing the following rights and obligations:

- No royalty or other fee imposed upon redistribution.
- Availability of the source code.
- Right to create modifications and derivative works.
- May require modified versions to be distributed as the original version plus patches.
- No discrimination against persons or groups.
- No discrimination against fields of endeavour.
- All rights granted must flow through to/with redistributed versions.
- The license applies to the program as a whole and each of its components.

- The license must not restrict other software, thus permitting the distribution of open source and closed source software together.

This definition clearly leaves room for a wide variety of licenses (see section 12). While the fundamental philosophy of the two movements are different, both FSF and OSI share the same space and cooperate on practical grounds like software development, efforts against proprietary software, software patents, and the like. As Richard Stallman says, the Free Software Movement and the Open Source Movement are two political parties in the same community.

But FOSS is more than a philosophy, it is also a software development approach that has resulted in the new and powerful software, of which some dominate the current software spectrum.

The changing concept and work approach that is used in open source software development was well described and analyzed by Erik Raymond in his book "The Cathedral and the Bazaar" (Raymond, 1998). The cathedral and bazaar analogies are used to contrast the FOSS development model with traditional software development methods.

Commercial software development is similar to the way cathedrals were built in ancient times. Small groups of skilled artisans carefully planned out the design in isolation and everything was built in a single effort. Once built, the cathedrals were complete and little further modification was made. Software was traditionally built in a similar fashion. Groups of programmers worked in isolation, with careful planning and management, until their work was completed and the program released to the world. Once released, the program was considered finished and limited work was subsequently done on it.

In contrast, FOSS development is more akin to a bazaar, which grows organically. Initial traders come, establish their structures, and begin business. Later traders come and establish their own structures, and the bazaar grows in what appears to be a very chaotic fashion. Traders are concerned primarily with building a minimally functional structure so that they can begin trading. Later additions are added as circumstances dictate. Likewise, FOSS development starts off highly unstructured. Developers release early minimally functional code to the general public and then modify their programs based on feedback. Other developers may come along and modify or build upon the existing code. Overtime, an entire operating system and suite of applications develops and evolves continuously.

The model of the bazaar is an interesting model for users and software industry in the LDC's. Since they have not been involved in the development of the 'software cathedrals' of modern times, their needs have not been addressed. Requests like translating e.g. Microsoft Office in local African languages (even the large ones like Swahili) land on deaf ears. In the bazaar model it becomes more easy to get the needs of the LDC's integrated, through collaborating in the development of new applications or forking[12] of existing applications.

## *7. Advantages and disadvantages of FOSS*

**KEYWORDS:**
FOSS, ADVANTAGES OF FOSS, DISADVANTAGES OF FOSS, NATIONAL ADVISORY COUNCIL OF ON INNOVATION SOUTH AFRICA, UK OFFICE OF GOVERNMENT COMMERCE, SUSTAINABILITY

The discussion about the advantages and disadvantages of FOSS is a difficult discussion since there are lack of objective information available. We will therefore list some of the advantages and disadvantages listed by others.

South Africa's National Advisory Council on Innovations summarizes the major benefits of FOSS and the adoption of open standards and software as promoted in the FOSS paradigm[13]:

- Reduced costs and less dependency on imported technology and skills
- Affordable software for individuals, enterprise and government
- Universal access through mass software rollout without costly licensing implications
- Access to government data without barrier of proprietary software and data formats
- Ability to customize software to local languages and cultures
- Lowered barriers to entry for software businesses
- Participation in global network of software development

Additional advantages that are identified the UK Office of Government Commerce (OCG, 2002) are:

- Supplier independence, limiting vendor lock-in
- Patches or updates become available quicker, which limits breakdowns and security risks

At the same time there are also limitations and drawbacks to the use of FOSS. The UK Office of Government Commerce identifies the following factors that my limit successful implementation:

- *Available support for FOSS.* In the past years support has been lacking a professional approach. In recent years this has improved now that large software companies like IBM, SUN and HP have started to join the FOSS movement.

- *Finding the appropriate software*: Since FOSS is not 'advertised' it can be very difficult to select the appropriate applications for the task it has to support. A more active approach is needed from the users.
- *Documentation*: The documentation that accompanies FOSS software application is often idiosyncratic and sometimes non-existent. FOSS developers are motivated towards the technical aspects of the application than towards the usability.
- *Limited best practices*: There are very little known and documented cases of large scale migration from commercial software to FOSS.
- *Hardware – software fit*: FOSS often lags behind concerning new hardware. This is caused by the fact the hardware manufacturers fail to release hardware specifications in time to the FOSS community.

The bazaar method of software development has been proven over time to have several advantages:

- *Reduced duplication of effort*: By releasing programs early and granting users the right to modify and redistribute the source code, FOSS developers reuse the work produced by compatriots. The economies of scale can be enormous. Instead of five software developers in 10 companies writing a single networking application, there is the potential for the combined efforts of 50 developers. The reduced duplication of effort allows FOSS development to scale to massive, unheard of levels involving thousands of developers around the world.
- *Building upon the work of others*: With the availability of existing source code to build on, development times are reduced. Many FOSS projects rely on software built by other projects to supply the functionality needed. For example, instead of writing their own cryptographic

　　　　code, the Apache web server project uses the OpenSSL project's implementation, thereby saving thousands of hours of coding and testing. Even in cases where source code cannot be directly integrated, the availability of existing source code allows developers to learn how another project has solved a similar problem.

- *Better quality control*: "Given enough eyeballs, all bugs are shallow" is an oft-cited quotation in the FOSS world. It means with enough qualified developers using the application and examining the source code, errors are spotted and fixed faster. Proprietary applications may accept error reports but because their users are denied access to the source code, users are limited to reporting symptoms. FOSS developers often find that users with access to the source code not only report problems but also pinpoint the exact cause and, in some cases, supply the fixes. This greatly reduces development and quality control time.
- *Reduced maintenance costs*: Maintenance of any software package can often equal or exceed the cost of initial software development. When a single organization has to maintain software, this can be an extremely expensive task. However, with the FOSS development model, maintenance costs can be shared among the thousands of potential users of a software application, reducing per organization costs. Likewise, enhancements can be made by the organization/individual with the best expertise in the matter, which results in a more efficient use of resources.

The advantages are alike for the developed and developing countries, but some have more weight in the LDC's. The most obvious aspect is the cost aspect, for FOSS users (individuals and organizations) pay no licensing fee. Cost reduction, especially recurrent costs, is increasingly important in Africa, to become less dependent on donor grants. The Total Cost of Ownership, is often mentioned to be higher for FOSS since more development time (with expensive developer salaries) is needed. In the LDC's where salaries are significantly lower, this may tip the scales to the other side.

However, the "openness" and flexibility of FOSS is more important when considering the situation at hand in Africa. FOSS can be customized and constantly revised to develop and change with the needs of the user. It is only now when ICT is implemented in the LDC's that the needs and requirements for the software is gradually discovered. Moreover, where propriety software is very hardware intensive, FOSS can be be modified to run on computers that are "obsolete". This will limit the need to replace hardware frequently.

Of all the advantages and disadvantages the open software development communities may prove the biggest advantage of FOSS in for the LDC's. Lecturers and trainers that are conversant with modern software technologies and tools are often hard to find in LDC's. This has a negative impact on the development of the technical capacity needed. Through the participation in bazaar like software development projects, implicit training in software development becomes available though other participants, that would otherwise not be accessible.

## *8. Is donated software also free software?*

**KEYWORDS:**
PROPRIETARY SOFTWARE, LDC'S, SOFTWARE AS GIFT, PRICE OF SOFTWARE, BASE OF THE PYRAMID (BOB), UBUNTU

Although it might be clear by now, FOSS is not the same as donated software. In recent years the software vendors have 'discovered' the potential of the LDC's. The International

Finance corporation of the World Bank group and the World Resources Institute (Hammond et al., 2007) estimate the market for ICT and ICT related services at the so-called base of the pyramid (BOP) on USD 51.4 billion and growing rapidly. An interesting figure and the large proprietary software producers and vendors are rapidly establishing emerging market divisions to tap into this enormous potential.

Well aware of the fact that the spending power of these economies is not yet strong enough to afford expensive software solutions, offering low cost or free versions of their proprietary and more expensive commercial software is considered a viable first step to bind these new markets to their companies. With success. Several countries in Africa have standardized their national database systems on proprietary software, universities have adopted proprietary tools to support the learning processes for their computer science students and recently we see the development that in some countries national computer literacy exams for secondary school students are only granted on the Microsoft platform. The decision to adopt the proprietary platforms and software is justified by idea that the software is donated by the vendor at a low cost or even free.

This notion of 'free' should not be considered the same as the notion 'free' of that is used for Free and Open Source Software. The donated software may not require (much) investment, but in all other aspects the software is not free. It cannot be shared with other members of the community, the user is not allowed to adapt the software to the local needs, and the costs may be low today, tomorrow the owner of the software may ask you to pay for its use. In other words, this can be a free gift that will come with huge future costs.

Where the donated/free software still uses the license to restrict the user from sharing and redistributing the software and limits the user from adapting the software to local conditions but thus getting back to the software producer, FOSS encourages this. This best illustrated with the text on the Ubuntu CD cover[14]:

*"Ubuntu is software libre. You are encouraged and legally entitled to copy, reinstall, modify and redistribute this CD for yourself and your friends"*

and

*"Ubuntu will always be free of charge, including enterprise releases and security updates"*

Until software donations are performed under these conditions, the 'free' will come with limitations and an expiration date.

## *9. What softwares are well-known free and open softwares – desktop?*

**KEYWORDS:**
FOSS, LINUX, LINUX DISTRIBUTIONS, PRODUCTIVITY SOFTWARE, USER SOFTWARE, BUSINESS SOFTWARE, SMALL BUSINESSES, SOFTWARE ALTERNATIVES

Software is an essential element in the operation of every computer, from PDA to notebook, from desktop to server. At a general level we identify two types of computer software: operating systems software and application software. We could introduce more complex classification of the different software layers, as the OSI model, but they are beyond the scope of this book.

Operating systems software is designed the make all the different hardware components in the computer, as well as all the peripherals, work together and to make it operate as an integrated machine. The operating system does interpret signals from keyboard and other input peripherals, allowing the user to input data, to process it in the central processing unit, store it temporarily or permanently on storage devices, and provide an output on output peripherals, as screen or printer.

Linux is considered the main operating systems software FOSS alternative. Linux is the runaway success of the Unix world. The term Linux is often used synonymous with the Linux distribution . The distribution is the Linux operating system software (kernel) bundled with application and/or server software. In some cases the distribution is a bundling of thousands of bigger and smaller applications. There is however only one Linux kernel and there are many Linux distributions. The best-know linux distributions[15] are listed in the table below. We have distinguished between fully FOSS distributions and partial FOSS, where FOSS is combined with some proprietary elements.

| **Fully FOSS** | | **Partial FOSS** | |
|---|---|---|---|
| Ubuntu | www.ubuntu.com | SuSE | www.suse.de |
| Slackware | www.slackware.org | Red Hat | www.redhat.com |
| Debian | www.debian.org | Mandriva | www.mandriva.com |

Table 2: The Major Linux Distributions with their Websites.

Application software is designed to support individual users or organizations in executing their tasks. Application software is used on top of the operation systems software. For most tasks that users perform on the desktop there are FOSS alternatives available. In the table below we have listed major tasks of the user and the most important FOSS alternatives that will support this task.

FOSS is often regarded as software that is designed for the Linux platform. However this is not necessarily the case.

Many of the FOSS applications work on the Linux operating system as well as on the Microsoft Windows operating system. In the table below we have therefore indicated the operating system the software will work on. We have selected, where possible, software that works on both Windows (indicated with W in the table) and Linux (L).[16]

| Task | Application | Website | Platform |
| --- | --- | --- | --- |
| Office productivity suite | Open Office | www.openoffice.org | W/L |
| Web browser | Firefox | www.mozilla.org | W/L |
| Email reader | Thunderbird | www.mozilla.org | W/L |
| Personal Information Management (calendars, tasks, addresses, emails etc) | Chandler Evolution Kontact | chandlerproject.org/ www.gnome.org www.kontact.org | W/L L L |
| Image Editing | GIMP | www.gimp.org | W/L |
| Desktop publishing | Scribus | www.scribus.net | L |
| Media player | VLC | www.videolan.org | W/L |
| Personal Database | Open Office Base | www.openoffice.org | W/L |

Table 3: The Main FOSS Alternatives for the User/Desktop Tasks.

Business software is often more expensive than user/desktop software and this poses a huge challenge for start up companies and small and medium enterprises (SME) in the LDC's. Although they are the driving force of many developing economies, the profits are small, financial institutions are reluctant to support investment for these organizations and therefore large investments in software are often not possible. However, in order to grow their businesses and expand abroad, the SME's will have to automate. FOSS provides a range of business applications that provide good alternatives for the expensive proprietary business software.

Below we present a list of some of the most important FOSS alternatives for common business tasks.

| Task | Application | Website | Platform |
|---|---|---|---|
| Customer Relationship Management | SugarCRM | www.sugarcrm.com | W/L |
| Document Management | Alfresco | www.alfresco.com | W/L |
| Financial Management | SQL Ledger<br>GNU Cash | www.sql-ledger.org<br>www.gnucash.org | W/L<br>L |
| Project Management | Open Project<br>Gantt Project | www.projity.com<br>www.ganttproject.org | W/L<br>W/L |
| Enterprise Resource Planning (including financial management) | CentricCRM<br>Adempiere | www.centriccrm.com<br>www.adempiere.com | W/L<br>W/L |
| Knowledge management | pbwiki | www.pbwiki.com | W/L |
| Web Content Management | Joomla<br>Drupal | www.joomla.com<br>www.drupal.com | W/L<br>W/L |
| Web Site Design | NVU<br>Quanta Plus | www.nvu.com<br>quanta.sourceforge.net | W/L<br>L |
| Database | MySQL<br>PostgreSQL | www.mysql.com<br>www.postgresql.org | W/L<br>W/L |

Table 4: The FOSS Alternatives for (Small) Business Tasks.

## *10. What softwares are well-known free and open softwares – server?*

**KEYWORDS:**
FOSS, SERVER SOFTWARE, SOFTWARE ALTERNATIVES, EMAIL SERVICES, DATABASE SERVICES, FILES SHARING SERVICES, WEB SERVICES

When using computers in a networked environment, the user is only confronted with a small proportion of all the software that is used. To connect and survive in a computer network the user is connected to one or more servers that contain information and software. For the user this software is mostly invisible and applications on the user side are used to navigate through the network without knowing the networking details. However, servers are recommended when more computers need to access the same data, and in many small and medium enterprises this is the case.

On the server-side, which is mostly operated by the network administrator or network operator, a lot of different applications and hardware are used to enable the major networking functions or services:

- **Email services**: In LDC's many small organizations use public email services like Yahoo! or Hotmail. When the organization becomes more professional services need to be set up a mailserver to send and receive mail.
- **Web services**: Many organizations acknowledge the importance of their presence on the World Wide Web (WWW) with a website with corporate information becomes more important. In order to do so, a web-server needs to be set up.
- **File Sharing services**: When working in a network with information and data on a central server, there is need for file sharing services.

- **Database services**: Getting information and storing information in the business is best done with databases. When an organization grows, central database systems will be introduced.

FOSS has a bigger impact on the server environment than it had in the user/desktop environment. Many system administrators find FOSS interesting since it offers alternatives that require or little or no investments (Upadhaya, 2007). Presently, most of the Internet Service Providers and Telecommunication providers in the LDC's use FOSS for their servers.

| Task | Application | Website |
| --- | --- | --- |
| Mail server | Postfix<br>Sendmail | www.postfix.org<br>www.sendmail.org |
| Web server | Apache | www.apache.org |
| Database server | MySQL<br>PostGres | www.mysql.com/<br>www.postgresql.org/ |
| File sharing server | Samba | us1.samba.org/samba/ |
| Content filtering server | SquidGuard | www.squid-cache.org |
| Security server | NMap | www.insecure.org/nmap |
| Anti-virus | ClamAV<br>Amavis | www.clamav.net<br>www.amavis.org |
| Spam filtering | SpamAssassin | www.spamassassin.org |

Table 5: FOSS Alternatives for the Server Environment.

The server environment is often a major hurdle for organizations in LDC's since there are limited experts available that can setup and manage a complex server environment with all the components above. In the FOSS world there are some excellent Linux distributions that offer all the applications that are needed to set up a server. Currently one of the best examples is SME-Server. SME-Server provide a distribution that installs out-of-the-box a webserver, a mailserver, a network file server, a firewall, content filtering and more. A relatively new direction for organizations in LDC is to use web-services like Google Apps.[17] This service allows organizations to host their email, webserver, and most of the other services above for virtually no costs. The server management is done by Google in an secure environment in the USA. This not really a FOSS solution, but very useful in an environment where limited qualified staff is available.

## 11. Who are the main stakeholders in the FOSS arena?

**KEYWORDS:**
FOSS, STAKEHOLDER ANALYSIS, SOFTWARE INDUSTRY, GOVERNMENT, DONOR COMMUNITY, LOCAL SOFTWARE INDUSTRY, CIVIL SOCIETY, LOCAL BUSINESS COMMUNITY, EDUCATIONAL INSTITUTIONS

In order to understand the FOSS we need to have an overview of the different players that participate in the community and the stakes that they have. The main stakeholders are listed below:

- **Software industry**: The key players in the FOSS arena are the software manufacturers, both producing and distributing proprietary software and FOSS. In recent years, the proprietary software industry has shown an increasing interest in the LDC's as a new sales frontier. Decision-makers and responsible government officials are approached in order to standardize on propriety software. Interesting 'free software' deals are offered. Unfortunately, the FOSS vendors have shown relatively little interest in the LDC market, with the exception of Ubuntu.
- **Governments**: Governments are the central players in the arena. The other stakeholders fight for their attention

in order to make them create the 'right' rules, regulations and laws. In the LDC's governments are mostly corrupt and therefore the outcomes of the decision making processes are unpredictable and not transparent (Laffont, 2005).
- **Donors**[18]: In the LDC's the power of the donor is mainly determined by the amount of funds they make available for development of the key issues in the country. Almost all donors invest in ICT as part of their approach, but there are only few donors that are specialized in explicitly devising ICT solutions for development. In general donors have good relations with decision-makers and government officials. Few donors have relations within the (local) ICT or software industry.
- **Local ICT industry**: The local ICT industry in LDC's is often young, immature and with a low level of organization. Individual businesses and entrepreneurs are fighting their way into a new market. Because of the short history of computing in LDC's these businesses are run by young people, that have recently graduated from local universities or expatriates that try to capitalize on the skill and knowledge advantage. In few countries the ICT industry has organized themselves in industry branch organizations that are able to put pressure on the government and decision-makers.
- **Local business community**: The local business communities increasingly depend on the ICT climate in a country. ICT is getting more and more important for their survival in the global economy and a good ICT infrastructure is a precondition for their international success. The local business community do have influence on the direction of the government policies, but only to a limited extent.

- **Civil society**: Like the local business community, civil society is aware that access to ICT and information plays a significant role in the country's development. They will try to influence government and decision-makers to improve regulations that promote access information and communication possibilities for all citizens. However, their influencing powers are limited.
- **Educational institutions**: The educational system provides the next generation computer users and ICT experts in a country. Most universities in the LDC's have a basic ICT infrastructure, train students to use computers and offer courses in Computer Science and sometimes in Information Systems. In an increasing number of secondary schools students have access to computer technology and some countries have made computer studies a compulsory subject for secondary school students. Governments set the guidelines for curricula for schools and play an important role in types of systems and platforms that are used.

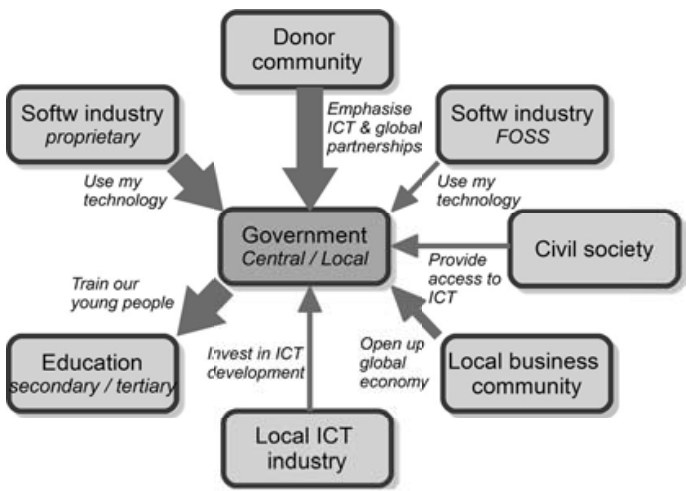

Figure 3: Stakeholders and their stakes in the FOSS Arena.

In the figure above we have displayed the stakeholders relationships. The arrows display the direction of the relationships, and the thicker the arrows are, the stronger the influential relationship is.

## 12. *What licenses are used for FOSS?*[19]

**KEYWORDS:**
FOSS LICENSES, OPEN SOURCE DEFINITION, GENERAL PUBLIC LICENSE (GPL), COPYRIGHT, OPEN SOURCE LICENSE, FREE SOFTWARE LICENSE, CLOSED SOURCE SOFTWARE

All software comes with licenses. The license protects the author of the software from others copying the software without his/her permission. Basically the license is an implementation of the basic copyright laws that have been in use for decennia in most countries around the world. This also implies that copyrights apply, even when they are not registered officially. When someone writes a small computer program for the school-bell to ring every 45 minutes for a period of 10 hours per day, but not on Sundays, the program is copyrighted simultaneously with its creation and is the sole property – barring any contractual abrogation of the copyright – of its creator. This counts for people in Europe, the USA, Asia and also for work that is done in most of the countries in Africa.

Open source licenses may be broadly categorized into the following types: (1) those that apply no restrictions on the distribution and (2) those that do apply such restrictions. This has resulted into two licensing paradigms: Free and Open Source Software (FOSS) and Proprietary and Closed Source Software (PCSS). Although both types of licenses are to protect the ownership of the software, they greatly differ in the extent to which they protect the rights to modify and redistribute and sell the product as well as the underlying software code.

The fundamental purpose of open source licensing is to deny anybody the right to exclusively exploit a work. Typically, in order to permit their works to reach a broad audience, and, incidentally, to make some sort of living from making works, creators are required to surrender all, or substantially all, of the rights granted by copyright to those entities that are capable of distributing and thereby exploiting that work.

Within the FOSS community, we identify two major trends in licensing: Open Source (OS) licenses and Free Software (FS) licenses.

FS licenses are the OS approach in its stronger form. FS licenses propagate indeed complete freedom to use the software's source code for any purpose and in any environment. The user of the packages released under FS licenses are granted complete access to the source code, as well as the right to all modification, to redistribute copies so that you can help your neighbor, and to improve the software and to release the improvements to the public so that the community can benefit. No constraints are allowed, and FS licenses in its strongest form, the GNU GPL license, are self-propagating, *id est* every modification to the source code of a package, which had originally been released under GPL, must be released under the same license. A package released under GPL can only evolve and be used in other packages, which are released under GPL themselves. The details for Free Software licenses are defined in the GNU Manifesto[20] and under this license high quality software has been produced since 1984.

The OS licenses are defined by the Open Source Definition. The Open Source Definition builds on the GNU Manifesto, but tries to provide credits to the originator of the software and to protect a product that is already in the market from misuse. At present there are more than 30 different licenses that are harbored by the Open Source Initiative. They differ from each other in the extend to which modification, redistribution and (re)selling of the software is allowed. Most important, packages released under some of the OS licenses, which still comply with the Open Source Definition, do not necessarily have to be released under the same license. Theoretically, a package released under CS license could then be built on top of another package released under OS license, even if the original OS licensed package has to be distributed along with the added CS components.[21]

Notwithstanding these secondary differences, FS and OS licenses are perfectly compatible, and FS licenses are indeed all Open Source Definition compliant, that is FS licenses are all also OS licenses, whilst the opposite is not true.[22]

Within the Closed Source Software community, it is normal practice that each software producer designs their own license that goes with the software. Large software companies like Microsoft and Oracle has specially designed user licenses, smaller organizations mostly work with standardized licenses to protect the intellectual property of their software. More information about the license that proprietary software producers use can mostly be found on their website. The information for the Microsoft products can be found at: www.microsoft.com/about/legal/useterms/

When a consumer purchases a piece of PCS software, say, Microsoft Excel, he or she acquires, along with the physical copy of the software and the manual (if there are such physical copies), the right to use the software for its intended purpose – in this case, as a spreadsheet program. By opening the plastic wrap on the box, the consumer becomes bound by the so-called "shrinkwrap license" under which s/he is bound not to copy the work (beyond the single copy made for her or his own use), not to make derivative works based on the work, and not to authorize anyone else to do either of these two things. The elimination of these three restrictions is the foundation of open source licensing.

Over the past decades several a growing number of licenses have been put forward to protect the products that are produced in the Open Source Arena. The most important licenses are:

- MIT
- BSD
- Apache, Versions 1.0 and 2.0
- Academic Free License (AFL)
- GNU General Public License (GPL)
- GNU Lesser General Public License (LGPL)
- Mozilla Public License (MPL)

How can one choose between the type of license required? First of all in practical and realistic terms, copyright issues lead their own life in LDC's. Most of the countries in Africa have a thriving illegal software market. Not only are illegal copies of most PCS software sold in markets and small roadside shops, there is also a lucrative business that installs illegal software and provides maintenance services on it.

The illegal use and distribution of PCS software is common practice in Africa (and large part of Asia) and there are good reasons for that. We will list the main reasons:

- There are limited outlets that sell legal copies of PCS software. One will have to search for 'official' outlets, while illegal ones are readily available.
- There are limited possibilities for local support – vendors of illegal software provide better services than the 'official' dealer. When a help-desk needs to be called for support, this is in most cases outside the country and therefore not affordable.

- Software is unreasonably expensive when related to income of people. For details, e.g. the comparison of License Fees and GDP Per Capita by Ghosh (2003).
- Finally, most users are not aware and education on the nature and implications of software licenses, both Open and Closed. This is ignorance is made worse since most computers are bought with pre-installed illegal software from 'official' hardware dealers.

Second: The choice of the correct licensing model is beyond the scope of this book. More information can be found at www.fsf.org/licensing/licenses and www.croftsoft.com/library/tutorials/opensource/

## *13. What is the essence of the GPL?*

**KEYWORDS:**
FOSS LICENSES, GENERAL PUBLIC LICENSE, GPL IN LDC'S, FREE SOFTWARE FOUNDATION

The GNU General Public License (GPL License or just GPL) is one of the foundation Open Source licenses and was created by the Free Software Foundation (FSF). Characteristic of the GPL is that it explicitly requires that derivative works be distributed under the terms of the GPL and also that derivative works may only be permitted to be distributed under the terms of the license.

The purpose of the GPL is explained in detail in the preamble that is attached to the license. The preamble clearly and concisely sets out the three main purposes of the GPL. The first, and by far the most important, is to keep software free, in the sense that it can be distributed and modified without additional permission of the licensor. This imposes a mirror-image restriction on the licensee: while the licensee has free access to the licensed work, the licensee must distribute any derivative works subject to the same limitations and restrictions as the licensed work. The second purpose of the GPL is to ensure that licensees are aware that software under the license is distributed "as is" and without warranty. The third purpose (which is really a variant of the first) is that the licensed software be free of restrictive patents: to the extent that a patent applies to the licensed software, it must be licensed in parallel with the code.[23]

The GPL is one of the most used software licenses in the FOSS world, but at the same time very suitable in the context of LDC's. It allows the free distribution of software without the violation of any copyright laws. It allows the local software industry to take up a piece of software in the public domain and start localizing or changing it. The skills to write software from scratch is mostly lacking and localization (language, currency, etc) of the software are mostly needed. The result of adaptation of the software will then again be available for other people who cannot afford or do not have the knowledge to make the changes. In this way, software developers build on the work of others while serving development goals.

## 14. What is Open Content?

**KEYWORDS:**
OPEN CONTENT, FOSS, FREE SOFTWARE MOVEMENT, OPEN CONTENT FOR LDC'S, GNU FREE DOCUMENTATION LICENSE

When talking about FOSS, one also needs to bring up the issue of Open Content. The developments in ICT are marked by the possibilities of greater dissemination of information and knowledge. Yet at the same time, stricter copyright laws that have been implemented over the last decennia (Lessing 2006, 2004) have created an invisible barrier to knowledge access and creativity in the information age. A number of scholars have used the metaphor of 'second enclosures' as a way of illustrating how the 'commons' of knowledge and culture are increasingly being fenced by the imposition of strict property protections on the intangible domain of intellectual property. It is in this context that 'Open Content' (and also FOSS) have emerged. These initiatives recognize that the future depends on proactively nurturing a vibrant 'commons' of knowledge and cultural resources.

Open Content derives philosophically from the Free Software movement and attempts to achieve for the world of general content what FOSS did for software. The word 'content' itself may sometimes be misleading as it refers to a whole range of subject matter, from music to movies and literature to learning materials.

The best known example of Open Content Development is Wikipedia.[24] Wikipedia is available under the GNU Free Documentation License. The encyclopedia's contents are written collaboratively by readers and are not subjected to any formal peer review. Readers can also edit the articles written by someone else. When using the material one does not have to pay for the use, but a reference to the source does need to be made.

A good example of an Open Content project directly aiming at the Developing World is the Global Textbook project.[25] The aim of the project is to develop, under the Creative Commons license, textbooks in the area of Information Systems, Computer Science and Business Studies that can be used by students in the developing world to overcome the prohibitive costs for traditional books. The project also provides the opportunity to use the basic texts but replace the examples with contextualized examples, i.e. Examples that reflect the situation in the country in which the book is used. This is important since business and technology contexts differ greatly in the developed and developing worlds.

## *15. What are the characteristics of Open Content licenses?*

**KEYWORDS:**
OPEN CONTENT LICENSES, CHARACTERISTICS OF OPEN CONTENT LICENSES, CREATIVE COMMONS LICENSE

Most Open Content licenses share a few common features that distinguish them from traditional copyright licenses. These can be understood in the following ways (Liang, 2007):

### Basis of the License/Validity of the License

While being a form of license that allows end users freedom, it is important to remember that the Open Content licenses, like Free Software licenses, are based on the author of a work having valid copyright. It is on the basis of this copyright and the exclusive rights that it grants him/her that the author can structure a license that allows him/her to impose the kinds of rights and obligations involved in using the work. Every Open Content license therefore asserts the copyright of the author and states that without a license from the author, any user using the work would be in violation of copyright.

### Rights Granted

The premise of an Open Content license is that, unlike most copyright licenses, which impose stringent conditions on the usage of the work, the Open Content licenses enable users to have certain freedoms by granting them rights. Some of these rights are usually common to all Open Content licenses, such as the right to copy the work and the right to distribute the work. Depending on the particular license, the user may also have the right to modify the work, create derivative works, perform the work, display the work and distribute the derivative works.

**Derivative Works**

Any work that is based on an original work created by you is a derivative work. The key difference between different kinds of Open Content licenses is the method that they adopt to deal with the question of derivative works. This issue is an inheritance from the licensing issues in the Free Software movement. The GNU GPL, for instance, makes it mandatory that any derivative work created from a work licensed under the GNU GPL must also be licensed under the GNU GPL. This is a means of ensuring that no one can create a derivative work from a free work which can then be licensed with restrictive terms and conditions. In other words, it ensures that a work that has been made available in the public domain cannot be taken outside of the public domain. On the other hand, you may have a license like the Berkeley Software Distribution (BSD) software license that may allow a person who creates a derivate work to license that derivative work under a proprietary or closed source license. This ability to control a derivative work through a license is perhaps the most important aspect of the Open Content licenses.

**Commercial/Non-Commercial Usage**

Another important aspect of Open Content licenses is the question of commercial/non-commercial usages. For instance, I may license a piece of video that I have made, but only as long as the user is using it for non-commercial purposes. On the other hand, a very liberal license may grant the person all rights, including the right to commercially exploit the work.

**Procedural Requirements Imposed**

Most Open Content licenses require a very strict adherence to procedures that have to be followed by the end-user if s/he wants to distribute the work, and this holds true even for derivative works. The licenses normally demand that a copy of the license accompanies the work,or the inclusion of some sign or symbol which indicates the nature of the license that the work is being distributed under, for instance,and information about where this license may be obtained. This procedure is critical to ensure that all the rights granted and all the obligations imposed under the license are also passed onto third parties who acquire the work.

**Appropriate Credits**

The next procedural requirement that has to be strictly followed is that there should be appropriate credits given to the author of the work. This procedure applies in two scenarios. In the first scenario, when the end user distributes the work to a third party, then s/he should ensure that the original author is duly acknowledged and credited. The procedure also applies when the end-user wants to modify the work or create a derivative work. Then, the derivative work should clearly mention the author of the original and also mention where the original can be found.

The best-known license in the Open Content domain is the Creative Commons license (www.creativecommons.org). The license is based on the philosophy that a large, vibrant public domain of information and content is a pre-requisite to sustained creativity, and there is a need to proactively enrich this public domain by creating a positive-rights copyright discourse. It does this by creating a set of licenses to enable Open Content and collaboration, as well as acting as a database of Open Content. Creative Commons also serves to educate the public about issues of copyright, freedom of speech and expression and the public domain.

The Creative Commons license comes in three main attributes:

1. **Attribution** – Gives permission to copy, distribute, display, and perform work and derivative works based upon it but only if credit is given.
2. **Noncommercial** – Gives permission to copy, distribute, display, and perform work and derivative works based upon it but for noncommercial purposes only.
3. **No Derivative Works** – Gives permission to copy, distribute, display, and perform only verbatim copies of work but not derivative works based upon it.
4. **Share Alike** – Gives permission to distribute derivative works only under a license identical to the license that governs the original work.

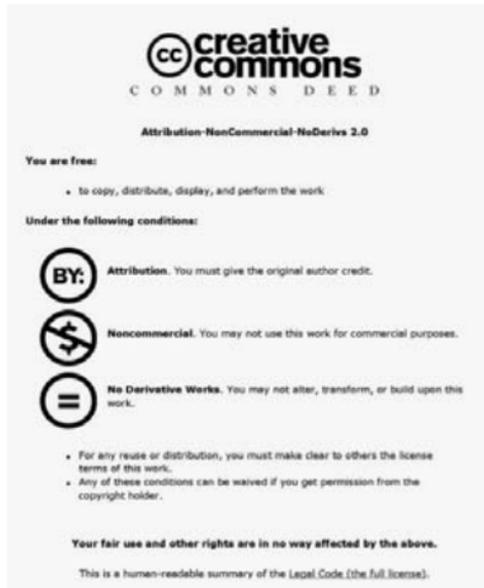

Figure 4: Creative Commons license as used by Eric von Hippel's book Democratizing Innovation.

## *16. Is FOSS only for LDC's?*

**KEYWORDS:**
FOSS FOR DEVELOPMENT (FOSS4D), DEVELOPMENT POTENTIAL, ECONOMIC DEVELOPMENT

We may get the impression that FOSS is something that is only applicable to the situation in LDC's. That is definitely not the case, however, the LDC's can benefit hugely from FOSS (Dravis, 2003). Weerawarana and Weeratunga (2004) conclude on the basis of case study research conducted mainly in Asia that careful exploitation of FOSS will enable LDC's to establish a global position in the IT driven knowledge economies of the future.

Ghosh and Schmidt (2006) list reasons why technologically advanced and LDC's alike should adopt FOSS as part of their ICT policies. In addition to the obvious cost-advantages, the study of FOSS developers and users communities demonstrate that the process of learning and adapting software enables the users to become 'creators of knowledge' rather than mere passive consumers of proprietary technologies. Through a system of 'informal apprenticeships' where the FOSS community takes care of the training of novices, local ICT competencies are being built. This new capacity, combined with the low entry barriers of FOSS, provides an excellent starting point for local business development. This link between FOSS and the rise of small ICT businesses is important given the tendency of proprietary vendors to ignore local needs, especially in developing and economically weak regions.

## *17. How can initiatives in FOSS be qualified?*

**KEYWORDS:**
FOSS LICENSES, MICRO LEVEL, MESO LEVEL, MACRO LEVEL, ICT POLICY

The FOSS arena is a complex world. It ranges from individual developers designing and writing programs that are offered to the public through a license like the GPL, to policy makers that promote national ICT policies to be changed to FOSS based ICT policies. This complexity makes it difficult to study and report on FOSS4D. Where to begin and what to address?

When we consider FOSS in the development context we have to concentrate on multiple levels in order to get a good understanding of the impact of the different initiatives. The implementation and the propagation of FOSS is performed on micro, meso and macro levels. At the *micro level* we like to think about individual users or small organizations (< 10 members) that decide for or against the use of FOSS. For example a user that prefers to use Open Office to make his or her texts, or a small NGO that decides to use a Linux mail server in stead of proprietary server software. At the *meso level* we consider organizations that take actions to integrate FOSS into their total software solution. These are organizations with a more complex organizational decision and governance structure and in most cases an already established (information/communication) technology infrastructure. In order to reach at a decision to implement FOSS, projects will need to have proposed and approved by the management of the organization. For example a university that likes to implement an Open Source Learning Management system like Moodle[26] will have to seek approval at different management levels in the university (faculty, senate, executive committee, ICT committee etc) before the project can be started. Finally, the *macro level* applies when government policies and actions are considered. At this level we will also find sector policies like educational policies that are proposed by government agencies or industry branch organizations. When a NGO representing e.g women's initiatives in a country publishes guidelines and recommendations to their members to use FOSS tools, we consider this to be an initiative at macro-level.

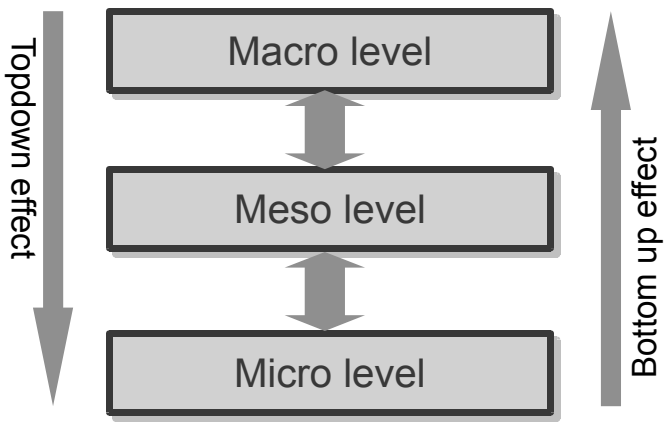

Figure 5: Framework for categorizing FOSS4D projects.

When considering the impact of projects, we also identify the possible effects the initiative can have on other organizations or individuals. The effects can take place in organizations and individuals within the same level, but it can also trickle down or up to the other levels. A school that has a very successful implementation of FOSS can serve as an example for other schools (meso level), but it can also make parents adopt FOSS on the computers at home (when applicable) or in the internet cafe around the corner, and ultimately, the experiences at one school may end up on the desk of a civil servant at the ministry of education who takes it as input for an ICT policy for the sector. These effects need to be promoted in the projects and programs that are initiated. At the same time we need to realize that not all projects have an impact as described above. Many individuals and small organizations that decide for the use of FOSS remain unnoticed.

In the three questions that will follow we illustrate the levels with examples that were found on the African continent. The examples will make the concepts more clear and provides some empirical evidence.

## *18. What are the key examples at a Macro level?*

**KEYWORDS:**
SUCCESS FACTORS MACRO LEVEL, BRAZIL, GITOC SOUTH-AFRICA, FOSSFA, SCHOOLNET NAMIBIA

Governments provide a huge potential for FOSS, not only as site for implementation for the software, but more importantly as propagators of the philosophy behind the FOSS movement.

Over the past years, a growing number of countries are starting to consider FOSS as a serious alternative (Hoe, 2006, Wong, 2004, Nicol, 2003). Brazil has been one of the countries that has actively pursued the FOSS model. It was in Brazil that the first law regarding the use of Open Source Software in the world was passed in March 2000. The country is one of the places where policies regarding adoption FOSS have been successful, notably in the states of Rio Grande do Sul and Pernambuco. Also, the Brazilian Navy has been using FOSS since 2002.[27]

In Africa, the South African government is the forefront player. In the wake of the developments, the South African government released a policy framework document in September 2002 by the open source work group of the Government Information Officers' Council (GITOC).[28] The GITOC Policy document (GITOC, 2002) recommends that government "explicitly" supports the adoption of open source software as part of its e-government strategy after a comprehensive study of the advantages and pitfalls of FOSS for government requirements. Next to adopting FOSS software GITOC also recommends that the government promotes the further development of FOSS in South Africa. There is an huge potential role for South Africa's SME industry in the production and implementation of FOSS as well as in setting up user training infrastructures. At the same time, the FOSS approach does represent a powerful opportunity for South African companies to bridge the technological gap, at an acceptable cost.

Some *success factors* need to considered in order to tap this potential:

1. *Implementation should produce value*: Value is related to economic value, i.e., the reduction of costs and saving of foreign currency; and social value, i.e., a wider access to information and computer training.
2. *Adequate capacity to implement, use and maintain*: There needs to be enough trained people to support and use the FOSS solution. Training users and developers needs to have a high priority.
3. *Policy support for an FOSS strategy*: Support for FOSS needs to expand to all key players at governmental level, departmental level, IT professionals and computer users in general.

Government's Department of Communication has already begun the move to Open Source by adopting Linux as their operating system. The South African government plans to save 3 billion Rands a year (approximately $338 million USD), increase spending on software that stays in their

country, and increase programming skills inside the country. South Africa reports that its small-scale introductions have already saved them 10 million Rands (approximately $1.1 million USD).

The government of Malawi has integrated the promotion of FOSS in the Malawi Nation ICT for Development Policy Document of September 2005:

> *"Advocate for the use of open source software as a viable alternative to proprietary software" (Section 3.3.2.1.1)*

Other countries are following.

Worldwide, similar moves are discussed by Taiwan, China, Viet Nam, the United Kingdom and Germany. Unfortunately, little governments in LDC's follow this direction.

An initiative with good potential that tries to bring together scattered FOSS society in order to get FOSS on political agenda is the Free and Open Source Foundation for Africa (FOSSFA).[29] The initiative started as the offspring of an ICT policy and civil society workshop in Addis Ababa, Ethiopia, in February 2003. During the workshop the participants agreed that FOSS is paramount to Africa's progress in the ICT arena. The mission of FOSSFA is therefore to promote the use and implementation of FOSS in Africa. Herewith it began to work on a coordinated approach to unite interested individual and to support open source development, distribution and integration. FOSSFA envisions a future in which governments and the private sector embrace open source software and enlist local experts in adapting and developing appropriate tools, applications and infrastructures for an African technology renaissance. They support South-to-South cooperation in which students from Ghana to Egypt and Kenya to Namibia develop softwares that are then adopted by software gurus in Nigeria, South Africa and Uganda in order to narrow the digital divide.

## *19. What are the key examples at Meso level?*

**KEYWORDS:**
MESO LEVEL, IICD, UGANDA MARTYRS UNIVERSITY, SCHOOLNET NAMIBIA

The International Institute for Communication and Development (IICD)[30], investigated the use of FOSS in organizations in three countries in Africa: Uganda, Tanzania and Burkina Faso (Bruggink, 2003). The objective of the research was to find out how, where and why organizations from all kind of sectors use FOSS, what problems can be observed and what opportunities for development are available. The findings of the research show that FOSS in Africa is being used, but it is not yet very widespread though there are huge differences between countries. FOSS is mostly found at the server side of Internet Service Providers (ISP's) and is sometimes used by government and educational institutions. This means that FOSS operating systems, mainly Linux and derivatives, web servers, email servers and files servers are found where the day-to-day computer users are not aware that they are actually using FOSS. Large and hierarchical organizations that have migrated completely from proprietary software to FOSS (server side and user side) were not noted in the report. Most of the organizations that are using FOSS are small organizations. When the three countries are compared, it is concluded that Ugandan organizations show most initiatives, while in Burkina Faso organizations do not show interest to move away from CSS.

The research of the IICD highlighted several reasons why organizations in Africa do not take up the challenge of FOSS. In the first place there are some false perceptions on FOSS. Many organizations believe indeed that FOSS is Linux only and that FOSS is user unfriendly and only suitable for the ICT specialist. Secondly, there is limited access to FOSS. Most of the FOSS is distributed through the Internet and with the limited and low bandwidth Internet connections, the access to FOSS is limited as a by product. Software companies, including FOSS companies, see little market potential in Africa (outside South Africa) and the availability of FOSS is low. This is also reflected in in the amount of resellers for FOSS. Finally, there is little expertise available to provide certified training and quality support for FOSS and eventually consultancy in migration processes.

A recent and interesting example of the introduction of FOSS at an organizational level is Uganda Martyrs University in Nkozi (Uganda). This migration is a role model for educational institutions on the African continent (Reijswoud, Mulo, 2006).

In 2002 Uganda Martyrs University embarked on a mission to be the first large organization in the region to completely migrate to FOSS. The main reasons for this decision were reduction of licensing costs and capacity building. At the start of the migration (August 2003) the university had about 250 desktop computers for students and staff, plus a variety of servers, connected in a campus wide Local Area Network. In 2002 the university started to migrate the server side (mail servers, Internet connection and file servers) of the network to FOSS. In the second phase of the project, which started June 2003, the university embarked on the migration the user side, after the university senate decided that all standard desktop computers of lecturing staff and students were to be equipped with a Linux operating system and FOSS applications like Open Office as a replacement for the popular Microsoft Office suite.

At present, next to the servers, all public (library) and student labs are migrated to FOSS. Staff computers have not been migrated, although a growing number uses FOSS applications in the Windows platform.

SchoolNet Namibia is another interesting example based in Africa. SchoolNet Namibia has developed a model for the empowerment of students through FOSS and the Internet which can act as a role model for the LDC's. SchoolNet Namibia started in February 2000 to empower youth through the Internet. Its main objectives were to provide sustainable low-cost ICT solutions to all Namibian schools. In this context it connected schools to the Internet, it did set up its own Internet Service Provider (ISP), it provides refurbished computers to schools, it implemented huge training programs for teachers and by now it connected 300 schools (in rural and urban areas) with 180,000 daily users, various libraries, teacher resource centers and non-government agencies, and it did set up computer laboratories in these schools and in many of the other resource centers. The schools and other centers use solely FOSS applications running on SuSE Linux and the schools are using the open source OpenLab application (which includes a bundle of educational content).

## *20. What are the key examples at Micro level?*

**KEYWORDS:**
MICRO LEVEL, AVOIR PROJECT, RULE PROJECT, EACOSS, BUSINESS SKILLS & DEVELOPMENT CENTRE

Most of the FOSS initiatives are small scale projects of individual people or small organizations.[31] A growing number of individuals throughout the LDC's is becoming aware of potential of FOSS from strategic point of view. This awareness results in smaller organization and individuals that start to develop or use FOSS.

To a limited extend, Open Source Software development projects have been launched in LDC's. On the African continent, most of the projects are situated in South Africa, for reasons connected to the presence of infrastructure. The African Virtual Open Initiatives and Resource (AVOIR) project, located at University of the Western Cape in South Africa is an interesting project which aim to develop FOSS capacity on selected universities in Africa.[32]

Outside South Africa, a project which is worth mentioning is the RULE[33] (Run Up to-date Linux Everywhere) project. The aim of the project is the creation of a very light Linux distribution for people that cannot afford modern computers systems. In order to achieve the goal, developers are modifying a standard Red Hat distribution, trying to allow the greatest real functionality with the smallest consumption of CPU and RAM resources. The new distribution is mainly intended to be for schools and other organizations in LDC's. At the present the RULE project provides a FOSS solution with GPL license that is able to transform 5 years old computer models (Pentium 75MHz, 16 MB RAM, 810 MB Hard disk) into useful machines again. Unfortunately, the project has stopped active development.

A recent and successful example in line with the RULE project is "One Laptop Per Child (OLPC)" project. The project uses FOSS to run on low-cost hardware.

The increasing interest for FOSS is also driving the emergence of FOSS specific organizations. In several countries of Africa, like Nigeria, Ghana, Uganda and South Africa, specialized software and consulting companies have started, whilst young people with a background in computing are embracing the FOSS approach and try to reform the accepted practice of buying (illegal) proprietary software. At present the market share of FOSS is still small and difficult for these specialized companies to grow, but when the benefits become clear and FOSS is implemented on a bigger scale, the capacity to implement the systems is ready.

That there is a demand for Linux and FOSS capacity is clear from the EACOSS[34] (East African Center for Open Source Software) project. The project, started in 2003 in Uganda and has trained a new generation of Linux and FOSS professionals in the region.

Business Skills & Development Centre, or BSDC is a South-African NGO was founded in 1987 by 3 women who saw the desperate need for young, black women to acquire job related skills in order to access meaningful employment in the business world. BSDC offers intensive courses of six-months in Business Administration, Office Skills and Entrepreneurship and includes typing, information technology, bookkeeping, office practice and business communication, business english, life skills and drama. Entrepreneurship is also taught in the form of theoretical and applied training. In fact BSDC is an incubator centre where students start their own small businesses with funds provided for by BSDC. The current intake is 50 students twice a year and 70% of the trainees find employment after the course. The computer laboratory is central to the training provided and its operational requirements are quite specific. However, licensing for proprietary software was found to be cost-prohibitive, and maintenance of individual, standalone PCs was also found to be too expensive.

The OpenOffice suite is used for teaching office computing while cost savings are achieved both through using free software and a low maintenance, terminal server environment (thin client solution). BSDC is in the process of moving their entire operation onto open source systems. The organization has just completed the first of four OS migration phases: the migration of the training environment to an open source terminal environment.

## *21. What lessons can be learned from the examples?*

**KEYWORDS:**
LESSONS LEARNED, RESISTANCE TO CHANGE, CHANGE IN MINDSET

The most important lessons that can be learned from the examples is that in spite of all the advantages, the actual use of FOSS is very limited in LDC's and that most projects are small. The migration of Uganda Martyrs University in Uganda has suffered enormous setbacks. Although the staff of the university is well aware of the advantages and value sustainable development, they resist to the change of their computer software. Similar resistance is also confirmed in smaller migrations of users with limited computer skills. In other words, knowing does not lead to doing.

We also observe that although LDC's are rapidly adopting ICT. However, using closed source software seems to be the norm. Since there are only a limited number of official software vendors, the origin of the software is dubious. Because of this origin, there is no official support for users and developers, but this does not seem to be a barrier. On the other hand, free community support for FOSS users is often presented as a key advantage. For the LDC's this advantage is not confirmed.

The limited number of LDC-based free and open source software development projects is very limited. Although some software has been localized, the does not seem to exist an practical software movement in the LDC's. The situation is worst in Africa. The challenges that are encountered in the AVOIR project and the slow progress that is made by the software developers outside South Africa is a reason for real concern.[35] The question needs to be asked whether we can expect software developers in the LDC's to contribute to the FOSS movement while they are struggling to make a living from ICT? It also seems that the FOSS approach needs more time to settle down. The change in mindset, the trust in other developers etc. need time to settle in in Africa

Finally, the lack of government attention for the use of FOSS is worrying. Where one would think that governments like to limit their public spending where viable alternatives are available, they decide to opt for expensive and proprietary solutions with annual recurrent licensing costs. Where governments have the possibility to build up an independent and open standards based ICT infrastructure and local industry, they seem to opt for strong vendor dependent and closed solutions. From a national development perspectives these decisions are hard to justify.

## 22. *What are the major hindrances for the introduction of FOSS in LDC's?*

#### KEYWORDS:
HINDRANCES, LACK OF INFORMATION, AVAILABILITY OF SOFTWARE, MISSING ROLE MODELS, LINUX USER GROUPS, UBUNTU, LAST MILE SOLUTION, EXTREMADURA

We identify three major factors that hinder the introduction of FOSS in LDC's: lack of information, availability of software and missing role models. We will consider these hindrances in more details below.

### Information

Access to information about advantages/disadvantages FOSS and alternatives to proprietary software is very limited. Most of this information is available and distributed through the internet, but the majority of the people in the LDC's still have limited access to this medium. When people have access they will only search for it when they are aware on the existence of FOSS.

Unfortunately, universities and schools pay very little attention to FOSS and Open Content. The large majority of schools and universities use (illegal) proprietary softwares for teaching and have little interest in alternatives. At the level of the teachers and lecturers there is too little knowledge about the FOSS in order to be a source of information for their students. In most LDC's the issue of copyright receives too limited attention to provide a start for a search for alternative solutions. The discussion of copyright laws could be a stepping stone for elaborating on FOSS and Open Content and the creation of awareness.

Some of the ICT-oriented donor community is informed about FOSS and will promote FOSS based solutions, however, the majority of the donor community promotes the use of ICT without addressing the FOSS issue. As mentioned before, they fail to take their guiding role to the level that they should.

Linux user groups (LUG) have emerged all over the developing world. They have become an important source of information for the Linux and FOSS communities. Because of their local focus they are able to serve the direct needs in the community, which are often different in the LDC's than in the global newsgroups. At the same time, these groups are mostly technology oriented and this may form a barrier for newcomers to join and participate.

**Software availability**

Like the information about FOSS, the software is also made available through the internet. There are hardly no physical distribution points for Free and Open Source Softwares except for the Ubuntu dissemination mechanisms (normal mail in some African countries and so-called toasters: a Linux operating system hooked up to a flat-screen where one can get copies of most open source operating systems and software for free.

This creates a huge barrier for the users of FOSS, since in most of the LDC's internet connections are slow, unstable and expensive. This makes the downloading of a complete distribution like SuSE or Fedora (1 Gb +) virtually impossible.[36]

Organizations like the East African Center of Open Source Software (EACOSS)[37] in Kampala tried to overcome this problem by using normal mail to bring the software in the country, then storing it in a public repository on their website and re-distributing it to users. This 'last-mile' solution is facilitated through scooter-taxis (boda boda) that take the software to the users for the costs of the CD-rom/DVD's and the boda boda fare. This initiative has been replicated in several other countries in Africa.

Ubuntu also recognized the issue and has made, from the beginning, their distribution available through mail. Users can order one or more copies of their software from the Ubuntu website.

**Missing role models**

A major hindrance to the growth of FOSS in the LDC's the lack of icons and iconic projects. In general the FOSS movement has only a limited number of people that can serve as examples for young entrepreneurs to look up to, contrary to the proprietary software movement where people like Bill Gates and Larry Allison spark imaginations of wealth and influence and for many people in the LDC's an escape from poverty. On the African continent there a virtually no other role models than Mark Shuttleworth of Ubuntu.[38] The leaders of the countries show no interest in FOSS and there are no businessmen that have made a fortune with the application of FOSS.

Similarly, there are very little large-scale projects that can serve as a model for young entrepreneurs. Projects like African Virtual Open Initiatives and Resources (AVOIR)[39], hosted by the university of the Western Cape in South Africa, that aim at the development of cutting-edge e-learning for the African continent has been able to attract the attention of academia, but has not been able to inspire the business community. Large projects like in Extremadura Spain, where the Ministry of Education, Science and Technology successfully initiated project to convert computer systems from proprietary systems to FOSS are not replicated in LDC's. The Extremadura project has been able to revive general prosperity and business activity in a poor region in Spain, and ultimately the quality of life in the region (Nah Soo Hoo, 2007, APC, 2007, Dravis, 2003).

The LDC's need some good examples of successful organizations that have succeeded with the use of FOSS. Existing projects will have to be more closely monitored and deserve more attention by the donor community. New projects will have to be reported more broadly.

## 23. What does it take to start with FOSS?

**KEYWORDS:**
USERS, TECHNICAL PERSONNEL, POWER USERS

To start implementing FOSS in LDC's requires above all a lot of courage and persistence. Making the decision to use FOSS is a decision that will involve continuous justification. Users and technical personnel will challenge it because it means for most of them a journey into the unknown. The most difficult people to convince are the ones that have just enough knowledge to use computers to meet their needs. They fear that the 'new system' will put them back in the position of learners, a position they have worked hard for to outgrow. New users and power users pose less problems. The new users have such a challenge ahead to master the new computer skills that they do not mind whether they get the skills on FOSS or proprietary applications. Moreover, most of them do not know the difference. Power users have enough skills and often curiosity that they adapt easily.

When deciding to use FOSS internet connectivity is essential, especially for the technical staff working on the project. Since most of them will not have the skills at hand to solve the problems they encounter, internet (users group and websites) will be their main sources of answers. As observed in the migration of Uganda Martyrs University (see above), relatively simple problems, like what filesystem is installed and what filesystem is best able to deal with power-cuts, can get technical experts and the project stuck.

Basic technical knowledge and skills are needed to provide a basis for understanding FOSS. We have observed that many of the so-called computer experts in LDC's lack basic understanding of hardware and software. Being trained in a 'click, drag and drop' environment did not prepare them for the more challenging problems and questions. Small scripts or minor alterations to software to provide a contextualized solutions is often already beyond their technical abilities. However, it is this knowledge that is needed to start to explore the full potential of FOSS.

As in every project, there needs to be a champion who drives the project as a figure head. In the LDC's, where hierarchical and generational relations still carry more weight, this needs to be a politically accepted figure. In most cases these people are hard to find and difficult to commit to the project.

## *24. Considering migrating to FOSS?*

**KEYWORDS:**
TOTAL FOSS MIGRATION, PARTIAL FOSS MIGRATION, PILOT MIGRATION, SERVER MIGRATION, DESKTOP MIGRATION, INDIVIDUAL MIGRATION, SKILLS

FOSS migrations can be distinguished in two types:

- Total FOSS migration: All software (operating system and applications) used on the computers (servers and user-workstations) in the organization is Free and Open Source.
- Partial FOSS migration: Some software used on the computers in the organization is Free and Open Source Software while other is proprietary.

Total migrations are very rare and in most of the cases the migration will aim at certain applications. In a similar line, organizations that only use proprietary software are rare. Most organizations use some FOSS applications (like Apache) on their servers.

When considering migrating, most organizations will start with a pilot migration. The actual goal of the pilot migration is not to have some computers in the organization use FOSS. The main goal of a pilot migration is answering the following question: How can we deploy across the organization with confidence? The key to a good pilot migration is that is includes all possible usage models that might also be included in an eventual migration (Almond et al., 2004).

Migrations are mostly initiated by the technical staff. Some system administrator or head of ICT department starts using FOSS applications, because s/he finds it interesting from a technical or cost-reduction perspective. A number of these people that are confronted with FOSS are dragged into the FOSS world (Individual migration). Through enthusiasm servers are migrated to Linux and other FOSS applications (Server migration) and in a small number of cases this extends to the productivity software on the users' desktops (Usage Area migration). Sometimes all software is migrated including a change of the operating system (Total migration). Uganda Martyrs University is one of the examples that went through these stages and started off on an almost total migration project (Reijswoud, Mulo, 2006). The four stages of FOSS migration are displayed in the figure below.

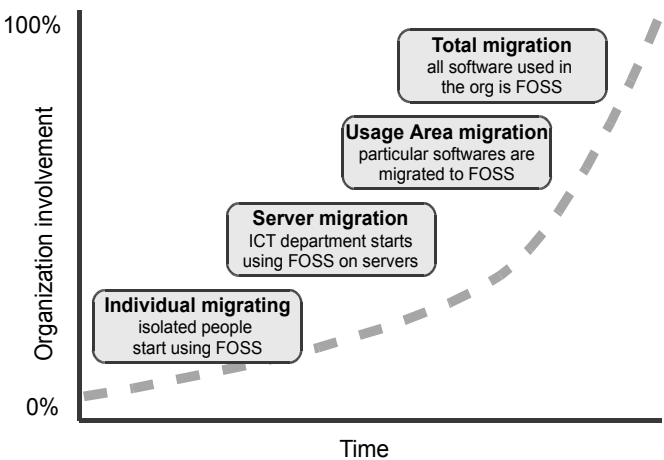

Figure 6: Four stages in FOSS migration.

A migration of an organization from a proprietary platform to an open software platform is considerably more difficult than starting to use FOSS from scratch. This is important for the people involved to realize. The initiators may be enthusiastic and the benefits clear to the management, but when it affects the users, they will be the main challenge. Computer users in the LDC's have very basic computer skills, but not withstanding how limited these skills are, they set them apart from the crowd. Moreover, these skills are often acquired through privately funded (expensive) computer training. The combination of the shallowness of their skills and value they present to them creates fear among the users that they will lose their acquired position. Strong resistance is the result. Only through extensive and continuous information and training sessions they will be willing to support the change. An additional sentiment was observed in the case of Uganda Martyrs University where the users felt that they were confronted by mediocre software. They replied to the people in charge of the migration: "If the software is so good, why are our colleagues in the West not using it?", is an argument hard to counter, especially when the donor promoting the use of FOSS makes a presentation with Microsoft PowerPoint and requests a .doc file when an Open Office file is sent.

## *25. Is there hope for FOSS in LDC's?*

**KEYWORDS:**
ADVANTAGES OF FOSS, ADVOCATING

Although empirical data on the use and impact of FOSS is still quite limited one can conclude that the penetration of FOSS in LDC's, especially in Africa is still low. Although there are no exact figures available, there seems to be a slight growth in the server segment, but hardly no growth in the user-desktop environment. This raises the question whether there is still hope for FOSS in LDC's.

The advantages of FOSS are clear and they are getting confirmed by the organizations and individuals adopting it. There is a reduction of costs, no vendor lock-in, unrestricted distribution of software and an increased understanding of computing at all levels involved. At the same time users feel isolated and different from their peers using mainstream software, and have fears that their skills in FOSS are less valuable on the job market.

The FOSS communities in the LDC's are fighting an uphill battle. Over the past years, the FOSS communities have achieved a relatively good level of organization in user groups, interest groups, training centers and some large conferences like Idlelo and AfricaSource. It was possible to make a lot of noise, but the question is: who is going to listen to them? There some areas like networking, system administration and internet hosting/website design where they find a willing ear. In other areas like productivity software for users, educational software and databases they receive little attention and we observe a growing penetration of proprietary software.

The FOSS advocates need to realize that examples need to be set. The advocates will have to show the people that FOSS solutions work instead talking about the advantages. When users and decision makers are confronted with a well-working FOSS computer environment, there is hope that more will decide to take the step of adopting it.

If the FOSS penetration in LDC's is to grow, all stakeholders will have to carefully consider their role and how to move to a more sustainable ICT infrastructure for LDC's. In conclusion we will therefore outline the challenges for the government of LDC's, the donor communities promoting the use of ICT, the educational sector that will train the new generation of users and developers, and the software industry in both the developing and developed world.

## *26. What are the challenges for governments in LDC's?*[40]

**KEYWORDS:**
SUCCESSES, PROMOTION OF FOSS, ICT POLICIES, ICT STRATEGIES, LOCAL SOFTWARE INDUSTRY

Successes of the FOSS movements in Brazil, South Africa, Extremadura in Spain and some cities in Europe clearly

underpin the importance of central and local governments in the promotion of FOSS. If FOSS is not embraced by government, there will be no changes at the meso levels.

Governments in the LDC's have to realize that they will have to build an ICT infrastructure that will, eventually, provide access to all citizens in the country, province, region or municipality. Vendor lock-in is highest where a significant investment in a proprietary technology is already in place. This is hardly the case in most developing countries where computerization is only beginning. So re-training and other transitional costs of moving from proprietary technology to a low-cost open source technology are much lower in LDC's. At the moment the donor communities in the developed world are willing to support, with financial means and knowledge, the initial stages of the ICT infrastructure buildup. However, this donor support will not last forever and the governments in the LDC's need to anticipate on this.

Central and local governments need to reconsider their ICT policies and strategies with a sustainability perspective in mind. In countries where the financial means are limited and not guaranteed, recurrent costs of the ICT infrastructure need to be a low as possible. Software licenses that need to be paid now, or in the future, do not fit into a sustainable policy when there are viable alternatives.

There is an opportunity for governments of LDC's. Governments of LDC's could make a start by adapting OSS for the public sector. The software eligible for FOSS alternatives can be categorized into four major groups:

- e-Government portals and service delivery systems
- Desktop office applications
- Server environments and networking
- Collaboration software

In order to accommodate a sustainable local software industry that can serve the country, the region and that it can even play a role in a global economy, governments will have to promote vendor independent and open solutions. Through the use of open standards, the local software industry will be able to offer services and solutions that provide the basis for a sustainable ICT infrastructure that allows growth and interconnection without be hindered by vendor controlled software standards.

Interesting sources of reference for governments in LDC's are:

1. The initiative of the Australian Government to develop A Guide to Open Source Software for Australian Government Agencies[41] which was released in 2005 with the intention to "provide Australian Government agencies with background information and processes to better understand, analyze, plan for and deploy open source software (OSS) solutions in appropriate situations".

2. The research conducted by the Berlecon Research which was financed by the European Commission' (IST programme). This research resulted in a series of reports such as Basics of Open Source Software Markets & Business Models, Motivations and Policy Implications.[42] It also presents the penetration of Open Source software in the EU showing that half of local government authorities already use at least some Open Source.

Finally, since internet access is crucial for capacity development in FOSS, the governments will have to create conditions for low-cost and wide-spread internet access.

## *27. What are the challenges for the donor community?*

**KEYWORDS:**
NEW DEVELOPMENT PARADIGMS, LOCAL ICT SECTOR, ROLE MODEL, OPEN STANDARD DOCUMENTS

The main challenge for the donor community is to start practicing what they preach.

Over the past years many large donors have published research confirming the potential of FOSS for development of a sustainable ICT infrastructure for the LDC's. Several donors have supported projects for the development and implementation of FOSS in LDC's.

The body of knowledge has become rich and vast. In spite of this research, little of the new projects seem to benefit. Still most of the computers that are used in donor projects are equipped with proprietary software and there is no coherent approach by all donors to guide beneficiaries in discovering the suitability of FOSS in their projects.

Too often the ICT issues in projects are dealt with by non-specialized program managers that have no or too little understanding of ICT to select appropriate solutions. The donor community will have to increase the number of ICT specialists and increase the level of ICT knowledge, including the understanding of FOSS, among their program managers. The role of ICT is getting too important for development. The appointment of ICT specialists is justified in all projects in which computer technology is applied.

The donor community should become more aware of the opportunity to become a role model in the use of FOSS. At present the donor community preaches the advantages of FOSS, but fail to adopt it themselves. Very few of the donors use FOSS application, like Linux, Open Office, Thunderbird etc. Many of them will not accept open standard documents like the OpenDocument Text files (.odt) or OpenDocument Spreadsheet files (.ods) and in this way are forcing their partners in LDC's to use proprietary software by demanding the use of .doc and .xls files. This behavior has a strong discouraging effect on new FOSS users in LDC's. The donor community will have to start to realize that change in the behavior in the LDC's starts with change of the behavior of the themselves.

To emphasize this point, donor will have to realize that

1. The opportunities for co-operation and participation in development projects by a community of users fit naturally in the new paradigms of development co-operation. The current development models emphasize ownership, knowledge sharing, Public Private Partnerships, collaboration and communities of practice. FOSS can be considered as a tool to support these new development paradigms: The new insights in development co-operation and FOSS are in that respect a perfect match.

2. The fact FOSS can contribute to economic development by supporting the development of the local ICT sector fits well in modern development cooperation. Due to the open and cooperative nature of FOSS it is easy for local programmers to get involved in adapting or developing software thereby not only creating opportunities for using ICT's as a tool in the traditional development sectors but also the development of new income-generation opportunities.

## *28. What are the challenges for education?*

**KEYWORDS:**
FOSS LABS, CURRICULUM, FOSS COMPETENCY CENTERS

Like the donor community, the educational world in the LDC's will have to reconsider their own position and behavior. In the developed world, the academic world has been a major driver in the promotion and developments in FOSS, the academia in the LDC's will also have to stand up and promote the use of FOSS for their own benefit as well as the benefit of their countries.

The traditional educational structure, starting from primary schools up through to the university level, can often be an excellent training ground for FOSS. There are a wide number of strategies in this sector, we will list some below:

- **Installation of FOSS labs**: This will limit the costs of the lab and will result in students that are open to FOSS.

- **A vendor neutral curriculum**: Make sure that the curricula do not contain vendor specific skills and knowledge.

- **Enforce the use of legal software in school/universities**: Management will have to prevent the use of illegal software by staff and students. This will make people aware of the costs and alternatives.

- **FOSS competency centers**: FOSS knowledge becomes essential for computer science students. Set up centers to build this capacity (and groom a new generation of FOSS professionals).

The transformation from proprietary software to FOSS will affect the curricula and will require the existing staff to acquire new computer skills. The educational system can promote this learning process and reward fast movers.

## *29. What can the software industry do?*

**KEYWORDS:**
AVAILABILITY OF FOSS, DISSEMINATION OF FOSS, CERTIFICATION PROGRAMMES

There is a major challenge for the FOSS industry to increase the emphasis on FOSS in the LDC's. Presently, the role of companies that are specialized in the development and distribution of FOSS is too limited to have a significant impact.

Although the international FOSS world is largely made up of individuals and small companies, there are also some large companies that can make a difference in the FOSS for development world. These companies should take a global responsibility for the development in LDC's and through efforts that concentrate along two lines:

1. **Improved availability of the software**: As noted, most software is distributed through the internet. Due to the lack of affordable internet connection the access to FOSS applications is low. In order to promote the access to FOSS a dissemination program will have to be set up. Local FOSS training centers like EACOSS and Linux User Groups could be used as point of distribution.

2. **Increased access to affordable certification programs**: Certification programs play an important role in LDC's where quality of education is not always guaranteed. Although some local ICT companies have tried to develop relationships with large distributors like Red Hat and SuSE, little of these efforts have materialized in affordable certification programs. When set up, the costs of certification are too high to be competitive with certification programs like for example MCSE.

> Also, in spite of huge effort from the FOSS community, the LPI program, although low cost, has not spread widely on the African continent.
> In order to promote the use of FOSS distributors and vendors should support the set up of low-cost certification programs to promote FOSS skills development in LDC's. The Cisco/UNDP program could serve as an example.

## *30. What is the research agenda for FOSS4D?*

**KEYWORDS:**
RESEARCH, FOSS4D, CAPACITY DEVELOPMENT, GOVERNMENT, CUSTOMIZED APPLICATIONS

Over the past years Open Source Software and Free Source Software have matured into a serious alternative when considering new software. The methods and the tools supporting software development processes in distributed environments like FOSS communities on the Internet, have been refined over the past years. As a result software products from the FOSS community have reached levels of reliability and security that allows them to compete with commercially developed software. In turn this gives an important impulse to the growth of the community.

Although most of the implementations of FOSS are still on the server side, user side adoption of FOSS grows now that friendly environments, high functionality and reliable alternatives for office applications become available. Governments, like Germany, the Netherlands, United Kingdom, and South Africa on the African continent, start to promote the use of FOSS.[43] Financial and moral support for development and use of FOSS alternatives increases awareness and acceptance.

FOSS initiatives in LDC's are still very limited. Africa is still in the phase of early adoption. Except for the South African government, governments in Sub Saharan Africa do not take a strong position in promoting the use of FOSS. This is may be partly due to fact that they are not well informed about the possibilities of FOSS, but it may also be caused by the fact that these countries have a low level of expertise in the ICT field. At present the skills levels needed for implementing and maintaining FOSS are perceived as higher.

The software development community in Africa is still in its infancy. University programs in software engineering are of relatively recent date, and the quality of the programs is low due to lack of facilities, lecturing materials and, most importantly, knowledgeable and dedicated lecturers. Training programs in the development of FOSS are not in place, which makes that African developers have to rely heavily on the expertise in other parts of the world. High bandwidth Internet access is therefore a precondition for success.

In spite of the low adoption, the FOSS paradigm provides advantages that are relevant within the African context. The most obvious advantage is the costing aspect. With increased licensing costs combined with high penalties for illegal use of proprietary software, FOSS provides a low cost alternative. Once the software is acquired, it can be used to automate a whole organization, small or large. Especially in large organization this can lead to a significant cost reduction. A different angle on the costing aspect is the fact that FOSS can easily be designed to run on 'obsolete' hardware, like the efforts in the RULE project. The financial situation of many countries in Sub-Saharan Africa does not allow large investments in new and modern hardware. Streamlined software can extend the life-span of computer hardware without compromising on functionality.

From a capacity development point of view, the openness of the program source code provides the software development community in Africa with an insight on near-commercial software development. African software developers can participate in the world-wide FOSS development community and improve their skills from this participation.

From a macro perspective a wide-spread adoption of FOSS may provide governments in Africa in the position to negotiate better conditions and improved functionality for the software they acquire. At present governments are the largest buyers of software products in Africa, but they have virtually no influence on the functionality of the products they purchase.

Finally, the flexibility of the FOSS makes it the perfect candidate for developing customized applications, which can keep into account peculiarities and specificity of the different local cultures. By adopting the FOSS paradigm organizations do not only reduce their costs, but also support a different perspective on intellectual property. If software is 'owned' by everyone, it is also owned by the people in the LDC's. This 'ownership' also provides the possibility to influence the direction of its development, and new, LDC-inspired features like the development of user interfaces in local languages, may be proposed.

There is still a long way to go, but the potential benefits are there at the end of the journey. Adoption of the FOSS paradigm needs to be encouraged in the LDC's, as it will represent a significant change in the technological relationship between the North and the South, developed and less/least developed countries, as we will no longer have to solely rely on the technical expertise of those in the First World. And this represents the first true step towards true sustainability.

On the basis of current situation we conclude with the formulation of a 5 point FOSS4D research agenda.

1. Get a clear understanding of the reasons why governments and decision-makers in the LDC's are not giving wide-spread support for FOSS and Open Standards.

2. Get a clear understanding of the reasons why such a small part of the international donor community actively promotes the use (donor and beneficiary sides) of FOSS and Open Standards in their projects.

3. Get a better understanding of the role open content lecturing material can play in the promotion and spread of FOSS and how these lecturing materials should be designed and distributed.

4. Research the possibilities to reduce software copyright infringements in LDC's by establishing educational programs and offering alternatives.

5. Research appealing role models that can be used for the promotion of FOSS in LDC's.

# LITERATURE AND SELECTED READINGS

Literature and selected readings 93

# ABOUT THE AUTHORS

**Victor van Reijswoud** finished his formal education with a Ph.D. in information systems from Delft University of Technology in the Netherlands. After an extensive career in academia and industry in Europe he got involved in ICT for Development as professor at Uganda Martyrs University in Uganda and Université Lumière de Bujumbura in Burundi. Through experience he observed the potential role that FOSS might play in development. He has initiated several FOSS migration projects in LDC's and has been acting as invited speaker at FOSS and ICT4D related conferences. Currently he is active as independent ICT4D advisor, researcher and professor at Divine Word University in Madang – Papua New Guinea. In these capacities he aims to share his experiences and build a more open world. Dr Van Reijswoud resides in Port Moresby.

**Arjan de Jager** studied Physics and Mathematics at the University in Utrecht in the Netherlands. After his studies he worked as lecturer Computer Science in the Netherlands and Zimbabwe. From 1998 to 2008 he worked as Country Manager for the International Institute for Communication and Development (IICD – www.iicd.org) in The Hague in the Netherlands. He has been working on ICT projects in Uganda, Tanzania, Mali and Zambia. Recently he joined the Centre for Expertise (HEC – www.hec.nl) as senior advisor responsible for ICT and Policy Development in the public sector.

# NOTES

1 There has been quite a lot of discussion and sometime intense debate about the label for Free and Open Source Software. Several labels have been put forward and are defended fiercely. We believe that this is an academic discussion and will provide little benefits for the users. We will use the term Free and Open Source Software (abbreviated to FOSS) through out this book, unless a specific aspect of FOSS needs to be emphasized.

2 Accessed January 4th 2008.

3 See for more information: www.gnu.org and www.fsf.org.

4 Fifty countries are currently designated by the United Nations as "least developed countries" (LDCs): Afghanistan, Angola, Bangladesh, Benin, Bhutan, Burkina Faso, Burundi, Cambodia, Cape Verde, Central African Republic, Chad, Comoros, Democratic Republic of the Congo, Djibouti, Equatorial Guinea, Eritrea, Ethiopia, Gambia, Guinea, Guinea-Bissau, Haiti, Kiribati, Lao People's Democratic Republic, Lesotho, Liberia, Madagascar, Malawi, Maldives, Mali, Mauritania, Mozambique, Myanmar, Nepal, Niger, Rwanda, Samoa, Sao Tome and Principe, Senegal, Sierra Leone, Solomon Islands, Somalia, Sudan, Timor-Leste, Togo, Tuvalu, Uganda, United Republic of Tanzania, Vanuatu, Yemen and Zambia. The list of LDCs is reviewed every three years by the Economic and Social Council (ECOSOC) in the light of recommendations by the Committee for Development Policy. (United Nations, Least Developed Countries Report 2007).

5 See: http://www.itu.int/wsis/tunis/newsroom/stats/

6 www.internetworldstats.com

7 www.developmentgateway.org

8 www.comminit.com

9 Figure on salaries of individual computer users are not known, but this remark is justified for the situation in Africa where the Gross Domestic Product (real) is US$ 354 (excluding South Africa).

10 We will not elaborate further on hardware requirements for the LDC context. Although important, this is outside the scope of the book.

11 For a more detailed explanation of why software needs to be free see: "Why Software Should Be Free", (http://www.fsf.org/philosophy/shouldbefree.html).

12 Forking in software development is like branching: Programmers take a copy of a program and start to develop a new program.

13 NACI January 2002 – www.naci.org.za/docs/opensource.html

14 Text as displayed on the CD cover of Ubuntu Version 6.06 LTS for your PC.

15 For a complete list of the all FOSS and non-FOSS Linux distributions see: www.distrowatch.com

16 We do not include Apple's OSX operating system, since we consider this a partial proprietary Unix variant and highly comparable with Linux.

17 www.google.com/a

18 With the term donor we denote all foreign agencies that providing or support in terms of knowledge or skills in LDC's. So this includes both funding agencies as well as implementing agencies.

19 This chapter is mainly based on St. Laurent, 2004.

20 For details see: www.fsf.org

21 Examples: Ximain or Mac OS X.

22 For further details on different FOSS licenses see: www.opensource.org

23 A given piece of code may be subject to both a copyright and a patent. In order for the GPL to function properly, both copyright and patent licenses must be subject to the terms of the GPL.

24 www.wikipedia.org

25 http://globaltext.org

26 www.moodle.org

27 http://www.pernambuco.com/tecnologia/arquivo/softlivre1.html

28 See for details and discussion about FOSS in South Africa: www.oss.gov.za

29 http://osfa.allafrica.com/

30 www.iicd.org

31 An interesting overview of micro FOSS projects is described by Na Soo Hoe in Breaking Barriers: The Potential of Free and Open Source Software for Sustainable Human Development.

32 For more information see the project website: http://avoir.uwc.ac.za

33 www.rule-project.org/en/

34 www.eacoss.org

35 One of the authors served as board member on the project and reports from own observations.

36 In Papua New Guinea where internet is paid by the megabyte would cost the download of Ubuntu 7.04 (697.9 Mb) around €200. This excludes the updates that have to be installed after the installation.

37 www.eacoss.org

38 His status is waning now that he is not longer living in South Africa.

39 http://avoir.uwc.ac.za/. See also: Na Soo Hoe, 2006.

40 For a more in-depth coverage of this issue, see: Wong, 2004.

41 See http://www.sourceit.gov.au/sourceit/oss

42 see http://www.berlecon.de/studien/downloads

43 Wong, 2004, EU Observer, Linux conquers Microsoft in Munich, 2003 (http://www.euobserver.com/index.phtml?aid=11435),

EU Observer, EU institutions test alternative to Microsoft (http://www.euobserver.com/index.phtml?aid=11261), Volkskrant, 21 augustus 2002, Computerbranch knokt om overheid.

# LIST OF KEYWORDS

(Some keywords appear duplicated in order to facilitate the search)

    not be considered an Adaptation (as defined above) for the purposes of this License.

c) **"Distribute"** means to make available to the public the original and copies of the Work or Adaptation, as appropriate, through sale or other transfer of ownership.

d) **"Licensor"** means the individual, individuals, entity or entities that offer(s) the Work under the terms of this License.

e) **"Original Author"** means, in the case of a literary or artistic work, the individual, individuals, entity or entities who created the Work or if no individual or entity can be identified, the publisher; and in addition (i) in the case of a performance the actors, singers, musicians, dancers, and other persons who act, sing, deliver, declaim, play in, interpret or otherwise perform literary or artistic works or expressions of folklore; (ii) in the case of a phonogram the producer being the person or legal entity who first fixes the sounds of a performance or other sounds; and, (iii) in the case of broadcasts, the organization that transmits the broadcast.

f) **"Work"** means the literary and/or artistic work offered under the terms of this License including without limitation any production in the literary, scientific and artistic domain, whatever may be the mode or form of its expression including digital form, such as a book, pamphlet and other writing; a lecture, address, sermon or other work of the same nature; a dramatic or dramatico-musical work; a choreographic work or entertainment in dumb show; a musical composition with or without words; a cinematographic work to which are assimilated works expressed by a process analogous to cinematography; a work of drawing, painting, architecture, sculpture, engraving or lithography; a photographic work to which are assimilated works expressed by a process analogous to photography; a work of applied art; an illustration, map, plan, sketch or three-dimensional work relative to geography, topography, architecture or science; a performance; a broadcast; a phonogram; a compilation of data to the extent it is protected as a copyrightable work; or a work performed by a variety or circus performer to the extent it is not otherwise considered a literary or artistic work.

g) **"You"** means an individual or entity exercising rights under this License who has not previously violated the terms of this License with respect to the Work, or who has received express permission from the Licensor to exercise rights under this License despite a previous violation.

h) **"Publicly Perform"** means to perform public recitations of the Work and to communicate to the public those public recitations, by any means or process, including by wire or wireless means or public digital performances; to make available to the public Works in such a way that members of the public may access these Works from a place and at a place individually chosen by them; to perform the Work to the public by any means or process and the communication to the public of the performances of the Work, including by public digital performance; to broadcast and rebroadcast the Work by any means including signs, sounds or images.

    i)    "**Reproduce**" means to make copies of the Work by any means including without limitation by sound or visual recordings and the right of fixation and reproducing fixations of the Work, including storage of a protected performance or phonogram in digital form or other electronic medium.

**2. Fair Dealing Rights.** Nothing in this License is intended to reduce, limit, or restrict any uses free from copyright or rights arising from limitations or exceptions that are provided for in connection with the copyright protection under copyright law or other applicable laws.

**3. License Grant.** Subject to the terms and conditions of this License, Licensor hereby grants You a worldwide, royalty-free, non-exclusive, perpetual (for the duration of the applicable copyright) license to exercise the rights in the Work as stated below:
   a) to Reproduce the Work, to incorporate the Work into one or more Collections, and to Reproduce the Work as incorporated in the Collections;
   b) to create and Reproduce Adaptations provided that any such Adaptation, including any translation in any medium, takes reasonable steps to clearly label, demarcate or otherwise identify that changes were made to the original Work. For example, a translation could be marked "The original work was translated from English to Spanish," or a modification could indicate "The original work has been modified.";
   c) to Distribute and Publicly Perform the Work including as incorporated in Collections; and,
   d) to Distribute and Publicly Perform Adaptations.

The above rights may be exercised in all media and formats whether now known or hereafter devised. The above rights include the right to make such modifications as are technically necessary to exercise the rights in other media and formats. Subject to Section 8(f), all rights not expressly granted by Licensor are hereby reserved, including but not limited to the rights set forth in Section 4(d).

**4. Restrictions.** The license granted in Section 3 above is expressly made subject to and limited by the following restrictions:
   a) You may Distribute or Publicly Perform the Work only under the terms of this License. You must include a copy of, or the Uniform Resource Identifier (URI) for, this License with every copy of the Work You Distribute or Publicly Perform. You may not offer or impose any terms on the Work that restrict the terms of this License or the ability of the recipient of the Work to exercise the rights granted to that recipient under the terms of the License. You may not sublicense the Work. You must keep intact all notices that refer to this License and to the disclaimer of warranties with every copy of the Work You Distribute or Publicly Perform. When You Distribute or Publicly Perform the Work, You may not impose any effective technological measures on the Work that restrict the ability of a recipient of the Work from You to exercise the rights granted to that recipient under the terms of the License. This Section 4(a) applies to the Work as incorporated in a Collection, but this does not require the Collection apart from the Work itself to be made subject to the terms of this License. If You create a Collection, upon notice from any Licensor You must, to the extent practicable, remove from the Collection any credit as required by Section

      4(c), as requested. If You create an Adaptation, upon notice from any Licensor You must, to the extent practicable, remove from the Adaptation any credit as required by Section 4(c), as requested.

b) You may not exercise any of the rights granted to You in Section 3 above in any manner that is primarily intended for or directed toward commercial advantage or private monetary compensation. The exchange of the Work for other copyrighted works by means of digital file-sharing or otherwise shall not be considered to be intended for or directed toward commercial advantage or private monetary compensation, provided there is no payment of any monetary compensation in connection with the exchange of copyrighted works.

c) If You Distribute, or Publicly Perform the Work or any Adaptations or Collections, You must, unless a request has been made pursuant to Section 4(a), keep intact all copyright notices for the Work and provide, reasonable to the medium or means You are utilizing: (i) the name of the Original Author (or pseudonym, if applicable) if supplied, and/or if the Original Author and/or Licensor designate another party or parties (e.g., a sponsor institute, publishing entity, journal) for attribution ("Attribution Parties") in Licensor's copyright notice, terms of service or by other reasonable means, the name of such party or parties; (ii) the title of the Work if supplied; (iii) to the extent reasonably practicable, the URI, if any, that Licensor specifies to be associated with the Work, unless such URI does not refer to the copyright notice or licensing information for the Work; and, (iv) consistent with Section 3(b), in the case of an Adaptation, a credit identifying the use of the Work in the Adaptation (e.g., "French translation of the Work by Original Author," or "Screenplay based on original Work by Original Author"). The credit required by this Section 4(c) may be implemented in any reasonable manner; provided, however, that in the case of a Adaptation or Collection, at a minimum such credit will appear, if a credit for all contributing authors of the Adaptation or Collection appears, then as part of these credits and in a manner at least as prominent as the credits for the other contributing authors. For the avoidance of doubt, You may only use the credit required by this Section for the purpose of attribution in the manner set out above and, by exercising Your rights under this License, You may not implicitly or explicitly assert or imply any connection with, sponsorship or endorsement by the Original Author, Licensor and/or Attribution Parties, as appropriate, of You or Your use of the Work, without the separate, express prior written permission of the Original Author, Licensor and/or Attribution Parties.

d) For the avoidance of doubt:
   i. i.Non-waivable Compulsory License Schemes. In those jurisdictions in which the right to collect royalties through any statutory or compulsory licensing scheme cannot be waived, the Licensor reserves the exclusive right to collect such royalties for any exercise by You of the rights granted under this License;
   ii. e.Waivable Compulsory License Schemes. In those jurisdictions in which the right to collect royalties through any statutory or compulsory licensing scheme can be waived, the Licensor reserves the exclusive right to collect such royalties for any exercise by You of the rights

      granted under this License if Your exercise of such rights is for a purpose or use which is otherwise than noncommercial as permitted under Section 4(b) and otherwise waives the right to collect royalties through any statutory or compulsory licensing scheme; and,

  iii. f.Voluntary License Schemes. The Licensor reserves the right to collect royalties, whether individually or, in the event that the Licensor is a member of a collecting society that administers voluntary licensing schemes, via that society, from any exercise by You of the rights granted under this License that is for a purpose or use which is otherwise than noncommercial as permitted under Section 4(c).

e) Except as otherwise agreed in writing by the Licensor or as may be otherwise permitted by applicable law, if You Reproduce, Distribute or Publicly Perform the Work either by itself or as part of any Adaptations or Collections, You must not distort, mutilate, modify or take other derogatory action in relation to the Work which would be prejudicial to the Original Author's honor or reputation. Licensor agrees that in those jurisdictions (e.g. Japan), in which any exercise of the right granted in Section 3(b) of this License (the right to make Adaptations) would be deemed to be a distortion, mutilation, modification or other derogatory action prejudicial to the Original Author's honor and reputation, the Licensor will waive or not assert, as appropriate, this Section, to the fullest extent permitted by the applicable national law, to enable You to reasonably exercise Your right under Section 3(b) of this License (right to make Adaptations) but not otherwise.

## 5. Representations, Warranties and Disclaimer

UNLESS OTHERWISE MUTUALLY AGREED TO BY THE PARTIES IN WRITING, LICENSOR OFFERS THE WORK AS-IS AND MAKES NO REPRESENTATIONS OR WARRANTIES OF ANY KIND CONCERNING THE WORK, EXPRESS, IMPLIED, STATUTORY OR OTHERWISE, INCLUDING, WITHOUT LIMITATION, WARRANTIES OF TITLE, MERCHANTIBILITY, FITNESS FOR A PARTICULAR PURPOSE, NONINFRINGEMENT, OR THE ABSENCE OF LATENT OR OTHER DEFECTS, ACCURACY, OR THE PRESENCE OF ABSENCE OF ERRORS, WHETHER OR NOT DISCOVERABLE. SOME JURISDICTIONS DO NOT ALLOW THE EXCLUSION OF IMPLIED WARRANTIES, SO SUCH EXCLUSION MAY NOT APPLY TO YOU.

**6. Limitation on Liability.** EXCEPT TO THE EXTENT REQUIRED BY APPLICABLE LAW, IN NO EVENT WILL LICENSOR BE LIABLE TO YOU ON ANY LEGAL THEORY FOR ANY SPECIAL, INCIDENTAL, CONSEQUENTIAL, PUNITIVE OR EXEMPLARY DAMAGES ARISING OUT OF THIS LICENSE OR THE USE OF THE WORK, EVEN IF LICENSOR HAS BEEN ADVISED OF THE POSSIBILITY OF SUCH DAMAGES.

## 7. Termination

a) This License and the rights granted hereunder will terminate automatically upon any breach by You of the terms of this License. Individuals or entities who have received Adaptations or Collections from You under this License, however, will not have their licenses terminated provided such individuals or entities remain in full compliance with those licenses. Sections 1, 2, 5, 6, 7, and 8 will survive any termination of this License.

b) Subject to the above terms and conditions, the license granted here is perpetual (for the duration of the applicable copyright in the Work). Notwithstanding the above, Licensor reserves the right to release the Work under different license terms or to stop distributing the Work at any time; provided, however that any such election will not serve to withdraw this License (or any other license that has been, or is required to be, granted under the terms of this License), and this License will continue in full force and effect unless terminated as stated above.

**8. Miscellaneous**

a) Each time You Distribute or Publicly Perform the Work or a Collection, the Licensor offers to the recipient a license to the Work on the same terms and conditions as the license granted to You under this License.

b) Each time You Distribute or Publicly Perform an Adaptation, Licensor offers to the recipient a license to the original Work on the same terms and conditions as the license granted to You under this License.

c) If any provision of this License is invalid or unenforceable under applicable law, it shall not affect the validity or enforceability of the remainder of the terms of this License, and without further action by the parties to this agreement, such provision shall be reformed to the minimum extent necessary to make such provision valid and enforceable.

d) No term or provision of this License shall be deemed waived and no breach consented to unless such waiver or consent shall be in writing and signed by the party to be charged with such waiver or consent.

e) This License constitutes the entire agreement between the parties with respect to the Work licensed here. There are no understandings, agreements or representations with respect to the Work not specified here. Licensor shall not be bound by any additional provisions that may appear in any communication from You. This License may not be modified without the mutual written agreement of the Licensor and You.

f) The rights granted under, and the subject matter referenced, in this License were drafted utilizing the terminology of the Berne Convention for the Protection of Literary and Artistic Works (as amended on September 28, 1979), the Rome Convention of 1961, the WIPO Copyright Treaty of 1996, the WIPO Performances and Phonograms Treaty of 1996 and the Universal Copyright Convention (as revised on July 24, 1971). These rights and subject matter take effect in the relevant jurisdiction in which the License terms are sought to be enforced according to the corresponding provisions of the implementation of those treaty provisions in the applicable national law. If the standard suite of rights granted under applicable copyright law includes additional rights not granted under this License, such additional rights are deemed to be included in the License; this License is not intended to restrict the license of any rights under applicable law.

**Creative Commons Notice**

Creative Commons is not a party to this License, and makes no warranty whatsoever in connection with the Work. Creative Commons will not be liable to You or any party on any legal theory for any damages whatsoever, including without limitation any general, special, incidental or consequential damages arising in connection to this license. Notwithstanding the foregoing two (2) sentences, if Creative Commons has expressly identified itself as the Licensor hereunder, it shall have all rights and obligations of Licensor.

Except for the limited purpose of indicating to the public that the Work is licensed under the CCPL, Creative Commons does not authorize the use by either party of the trademark "Creative Commons" or any related trademark or logo of Creative Commons without the prior written consent of Creative Commons. Any permitted use will be in compliance with Creative Commons' then-current trademark usage guidelines, as may be published on its website or otherwise made available upon request from time to time. For the avoidance of doubt, this trademark restriction does not form part of the License.

Creative Commons may be contacted at http://creativecommons.org/.

**Publishing studies** series

Development organizations and International Non-Governmental Organizations have been emphasizing the high potential of Free and Open Source Software for the Less Developed Countries. Cost reduction, less vendor dependency and increased potential for local capacity development have been their main arguments. In spite of its advantages, Free and Open Source Software is not widely adopted at the African continent. In this book the authors will explore the grounds on with these expectations are based. Where do they come from and is there evidence to support these expectations? Over the past years several projects have been initiated and some good results have been achieved, but at the same time many challenges were encountered. What lessons can be drawn from these experiences and do these experiences contain enough evidence to support the high expectations? Several projects and their achievements will be considered. In the final part of the book the future of Free and Open Source Software for Development will be explored. Special attention is given to the African continent since here challenges are highest. What is the role of Free and open Source Software for Development and how do we need to position and explore the potential? What are the threats? The book aims at professionals that are engaged in the design and implementation of ICT for Development (ICT4D) projects and want to improve their understanding of the role Free and Open Source Software can play.